\documentclass[peerreview]{IEEEtran}

\usepackage[square,numbers,sort&compress]{natbib}
\usepackage{amsthm}
\usepackage{amsmath}
\usepackage{mathrsfs}
\usepackage{amssymb} 
\usepackage{bbm,dsfont}   
\usepackage{enumerate} 

\usepackage[table]{xcolor}
\usepackage{multirow}
\usepackage{tikz}
\usetikzlibrary{calc}
\usepackage{pifont}%

\usepackage{physics}

\usepackage{xcolor}

\usepackage{hyperref} %
\usepackage%
{hyperref} %
\hypersetup{
    colorlinks,
    linkcolor={blue!80!black},%
    citecolor={green!30!black},
    urlcolor={blue!80!black}
}

\usepackage{pgf}
\usepackage{tikz}

\newcommand\blfootnote[1]{%
  \begingroup
  \renewcommand\thefootnote{}\footnote{#1}%
  \addtocounter{footnote}{-1}%
  \endgroup
}

\usepackage{comment}

\newcommand{\prob}[1]{\Pr\left(#1\right)}
\newcommand{\given}{\mid}
\newcommand{\cprob}[2]{\Pr\left(#1\given #2\right)}

\renewcommand{\mathbbm}[1]{\text{\usefont{U}{bbm}{m}{n}#1}} 
\newcommand{\eps}{\varepsilon}

\renewcommand{\trace}{\mathrm{Tr}}

\newcommand{\identity}{\mathbbm{1}}

\newcommand{\cf}{\emph{cf.} }

\newcommand{\hm}{\hat{m}}
\newcommand{\htm}{\hat{m}}

\newcommand{\hM}{\widehat{M}}

\newcommand{\Aset}{\mathcal{A}}

\newcommand{\Fset}{\mathcal{F}}

\newcommand{\Hset}{\mathcal{H}}

\newcommand{\Lset}{\mathcal{L}}
\newcommand{\Mset}{\mathcal{M}}
\newcommand{\Pset}{\mathcal{P}}

\newcommand{\Vset}{\mathcal{V}}
\newcommand{\Qset}{\mathcal{Q}}

\newcommand{\Xset}{\mathcal{X}}
\newcommand{\Yset}{\mathcal{Y}}
\newcommand{\Zset}{\mathcal{Z}}

\newcommand{\markovC}[1]{%
\begin{tikzpicture}[#1]%
\draw (0,0.3ex) -- (1ex,0.3ex);%
\draw (0.5ex,0.3ex) circle (0.2ex);
\draw[white] (0.2ex,0) -- (0.5ex,0);%
\end{tikzpicture}%
}
\newcommand{\Cbar}{\markovC{scale=2}}

\theoremstyle{remark}	\newtheorem{theorem}{Theorem}
\theoremstyle{remark}	\newtheorem{lemma}[theorem]{Lemma}
\theoremstyle{remark}	\newtheorem{corollary}[theorem]{Corollary}
\theoremstyle{remark}	
\theoremstyle{remark} \newtheorem{definition}{Definition}
\theoremstyle{remark} \newtheorem{remark}{Remark}
\theoremstyle{remark} \newtheorem{example}{Example}

\theoremstyle{remark}	\newtheorem*{discussion*}{Discussion}

\newcommand{\code}{\mathscr{C}}															

\newcommand{\tset}{\Aset_{\delta}^{(n)}}													

\newcommand{\opC}{\mathcal{C}}

\title{The Multiple-Access Channel with Entangled Transmitters}

\author{\IEEEauthorblockN{Uzi Pereg,~\IEEEmembership{Member, IEEE}, Christian Deppe,~\IEEEmembership{Member, IEEE}, and 
Holger Boche,~\IEEEmembership{Fellow, IEEE}}
}

\begin{document}
\maketitle

{}

\begin{abstract}
Communication over a classical multiple-access channel (MAC) with  entanglement resources is considered,
whereby two transmitters share entanglement resources a priori before communication begins.
Leditzky et al. (2020)  presented an example of a classical MAC, defined in terms of a pseudo telepathy game, such that the sum rate with entangled transmitters is strictly higher than the best achievable sum rate without such resources. 
Here, we establish inner and outer bounds on 
the capacity region for the \emph{general} MAC with entangled transmitters, and show that the previous result can be obtained as a special case. 
It has long been known %
that the capacity region of the classical MAC
under a message-average error criterion can be strictly larger than with a maximal error criterion (Dueck, 1978).
We observe that given entanglement resources, the regions coincide.
Furthermore, we address the combined setting of entanglement resources and conferencing, where  
the transmitters can also communicate with each other over rate-limited links. Using superdense coding, entanglement  can double the conferencing rate.
\end{abstract}


\blfootnote{%
Uzi Pereg is with the Faculty of Electrical and Computer Engineering  and the Helen Diller Quantum Center, 
Technion -- Israel Institute of Technology, 3200003 Haifa, Israel (email: uzipereg@technion.ac.il).
Christian Deppe is with the Institute for Communications Technology,
Technical University of Braunschweig, 38106 Braunschweig, Germany
(email: christian.deppe@tu-braunschweig.de).
Holger Boche is with 
the Institute of Theoretical Information Technology, 
Technical University of Munich, 80290 Munich, Germany,
BMBF Research Hub: 6G-Life,
the Munich Center for Quantum Science and Technology (MCQST), Schellingstr. 4, 80799
Munich, Germany, and
the Munich Quantum Valley (MQV), 
Hans-Kopfermann-Str. 1,
85748 Garching, Germany
(email: boche@tum.de).}

\blfootnote{
This paper was presented in part at  the 2023  IEEE Global Communications Conference (GLOBECOM 2023), Kuala Lumpur, Malaysia, December 4-8, 2023.
The results will also be presented in part at the BMBF Congress on Security in the Information Society as part of the plenary talk entitled “Unleash the true power of quantum communication for security, resilience and technological sovereignty”
on March 7-19, 2025 in Berlin, Germany.
}

\maketitle

\setcounter{page}{1}

\section{Introduction}
Cooperation in quantum communication networks has  gained considerable attention recently driven by both experimental advancements and theoretical discoveries
\cite{vanLoockAltBecherBensonBocheDeppe:20p,Pereg:22p,PeregDeppeBoche:23p1,ZlotnickBashPereg:24p,BFSDBFJ:20b}. 
The multiple-access channel (MAC) is a fundamental model for network communication.
Quantum resources in communication over the MAC are considered in the literature in various settings. %
Winter \cite{Winter:01p}  derived a regularized characterization for the classical capacity region of the quantum MAC. %
Furthermore, 
the authors of the present paper \cite{PeregDeppeBoche:22p} considered the quantum MAC with cribbing encoders, whereby Transmitter 2 has access to (part of) the environment of Transmitter 1.
 Boche and N\"otzel \cite{BocheNoetzel:14p} studied the cooperation setting of a classical-quantum %
MAC with conferencing encoders, where the encoders exchange messages between them in a constant rate (see also \cite{DiadamoBoche:19a}).
 Hsieh et al.  \cite{HsiehDevetakWinter:08p} and Shi et al.  \cite{ShiHsiehGuhaZhangZhuang:21p} addressed the model where each transmitter shares entanglement resources with the receiver independently.

Leditzky et al.  \cite{LeditzkyAlhejjiLevinSmith:20p}  
presented an example of a classical MAC
such that  sharing entanglement between transmitters strictly increases the achievable sum rate (see also \cite{Notzel:20p,NotzelDiAdamo:20c,YunRaiBae:23a}).
Recently, the classical upper bound 
has been improved showing that the sum rate increases from at most
$3.02$ to $3.17$ bits per transmission \cite{SeshadriLeditzkySiddhuSmith:22c}. 
The channel construction in \cite{LeditzkyAlhejjiLevinSmith:20p} is based on  a pseudo-telepathy game \cite{BrassardBroadbentTapp:05p} where quantum strategies guarantee a certain win and outperform classical strategies.
Additional observations are developed in 
\cite{DoolittleChitambarLeditzky:22c} as well.
 Fawzi and Ferm\'e \cite{FawziFerme:22c} have also  established  separation in a more fundamental example, the binary adder channel, when the transmitters are provided with non-singaling correlations  \cite{QuekShor:17p}.
By analyzing the zero-error capacity region, it was shown that the sum rate increases from $1.5$ without correlation resources, to $1.5425$ bits per transmission in this case \cite{FawziFerme:22c}. 
We   have shown that the dual property does not hold for the broadcast channel, i.e., entanglement between receivers does not increase achievable rates \cite{PeregDeppeBoche:21p2}. %

The potential benefits of the sixth generation of cellular networks (6G) are significant, with anticipated improvements in latency, resilience, computation power, and trustworthiness for future communication systems, such as the tactile internet
\cite{FettwisBoche:21m},
which involves not only data transfer but also the control of physical and virtual objects.
Resilience plays a very important role in achieving trustworthiness. 
Quantum resources are expected to play a crucial role in achieving these gains, as highlighted by \cite{DangAminShihadaAlouini:20p} and \cite{FettwisBoche:21m}. By enabling cooperation between trusted hardware and software components, future communication systems could isolate untrusted components and substantially reduce the attack surface of the communication system  \cite{FettwisBoche:21m,FettwisBoche:22p}. In particular, cooperation in the form of conferencing encoders can significantly increase the resilience of the communication system to jamming attacks. In \cite{Z5}, MACs were investigated for which without conferencing no communication is possible in the case of adversely jamming attacks, but even with arbitrarily small positive conferencing rates reliable communication between the sender and the receiver is guaranteed. Conferencing can thus prevent denial-of-service attacks for these channels. For a complete characterization of these scenarios, see \cite{Z5, Z6}. Quantum resources and cooperation also offer additional advantages, such as improved performance gains for communication tasks and the reduction of the attack surface, making them highly promising for 6G networks, as discussed in \cite{TKWIBD:20p} and \cite{FitzekBoche:21p2}. Investigating communication with cooperation in the form of entanglement resources could lead to more efficient protocols for future applications, making this an interesting avenue for further research. Cooperation and entanglement are important techniques to meet the very demanding requirements of 6G and future communication systems. The potential of these techniques for 6G has not really been explored in the literature until now. This paper aims to initialize this research direction. We only consider senders and receivers that are trustworthy, i.e. that adhere to the specified communication protocol. Especially in complexity theory, other interactive communication scenarios have also been examined with regard to their complexity behavior. This research direction is also interesting for communication with manipulative communication parties and should also be investigated for 6G applications in the future.

Here, %
we consider
communication over a two-user classical MAC with  entanglement resources shared
between the transmitters, a priori before communication begins.
We establish inner and outer bounds on the capacity region for the \emph{general} MAC with entangled transmitters (see Figure~\ref{fig:MentangledTx}), and show that the  result by Leditzky et al.  \cite{LeditzkyAlhejjiLevinSmith:20p} can be obtained as a special case. 
We also point out the following change of behavior.
In general, achievable communication rates may also depend on the error criterion.
Without entanglement resources, %
Dueck \cite{Dueck:78p} showed that
the relaxation of a message-average error criterion can lead to strictly higher achievable rates, when compared with a maximal error criterion.
Here, however, we show that the capacity region with entangled transmitters remains the same, whether we consider a message-average or a maximal error criterion. Furthermore, we consider the combined setting of entanglement resources together with  conferencing between the transmitters (see Figures \ref{fig:MentangledTx_Conferencing} and \ref{fig:MentangledTx_Conferencing_Q}).
In this setting,
in addition to pre-shared entanglement resources,  %
the transmitters can now communicate with each other over either classical or quantum noise-free links at a  limited rate. 
In the classical conferencing setting, we derive inner and outer bounds on the capacity region of the MAC with entangled transmitters. In the quantum conferencing setting,
we establish an achievable region and show that entanglement  can double the conferencing rate through superdense coding.

The remainder of this paper is organized as follows.
Section~\ref{Section:Related_Work} reviews related work on the MAC with classical correlation between the transmitters. 
Section~\ref{Section:Preliminaries} includes  basic definitions and the model description.
In Section~\ref{Section:Main_Results}, we present our main results on the classical MAC with entangled transmitters, where the entanglement resources  can be either unlimited (Section~\ref{Subsection:Main_Result}), or at a limited rate (\ref{Subsection:Rate_Limit_EA}).
We examine the example by Leditzky et al. 
\cite{LeditzkyAlhejjiLevinSmith:20p} in Section~\ref{Subsection:Magic_Example}, and consider the effect of classical and quantum conferencing on the capacity region in Section \ref{Section:Main_Results_Conferencing}.
 We prove the main results in
 Sections \ref{app:CardCl}-\ref{app:etMAC_Conferencing}.
In Section \ref{app:CardCl},
we show that our formulas are exhausted with finite-cardinality auxiliary variables. In Sections \ref{app:etMAC_In} and \ref{app:etMAC_Out}, we prove the capacity theorems for entangled transmitters, and in Section~\ref{app:etMAC_Conferencing}, for both entangled and conferencing transmitters.
In Section~\ref{Section:Discussion}, we summarize and discuss open directions, and
in the appendix, we show  that the capacity regions associated with the average and maximal error criteria are identical.

\section{Related Work
}
\label{Section:Related_Work}

We briefly review the related work on the multiple-access channel (MAC) with classically correlated transmitters. 
In communication, shared entanglement is often thought of as the quantum version of common randomness (CR), which is a random key that is shared between the two parties.
Indeed, generating CR is one of the most fundamental uses of entanglement.
The MAC setting provides yet another demonstration of how
%
entanglement is much  stronger than the classical CR correlation. We will come back to this point later on (see Remark~\ref{Remark:Common_Randomness}).

\subsection{Basic Definitions for Classical Systems}
\label{Subsection:Classical_Notation}
 We use the following notation conventions for classical systems. %
Script letters $\Xset,\Yset,...$ are used for finite sets.
Lowercase letters $x,y,\ldots$  represent constants and values of classical random variables, and uppercase letters $X,Y,\ldots$ represent classical random variables.   We use $x^r=\left(x[i] \right)_{i=1}^r$ to denote  a sequence of letters from $\Xset$, where $r$ is a positive integer. %
 The distribution of a discrete  random variable $X$ is specified by a probability mass function (pmf) 
	$p_X(x)$ over a finite set $\Xset$.
	The respective i.i.d. distribution is denoted by $p_X^n$, namely
 $p_X^n(x^n)=\prod_{i=1}^n p_X(x[i])$.	%
 A random sequence $X^n$ and its distribution $p_{X^n}(x^n)$ are defined accordingly. 
The %
typical set $\tset(p_X)$ is the set of sequences $x^n\in\Xset^n$ such that $\left| p_X(a) -\frac{1}{n}N(a|x^n)%
\right|\leq \delta \cdot p_X(a)$, 
for every $a\in\Xset$, where $N(a|x^n)$ is the number of occurrences of the letter $a\in\Xset$ in the sequence $x^n$.
The definition is extended to a joint distribution $p_{X,Y}$ in a natural manner.

%

%
%
A discrete memoryless multiple-access channel (MAC) $(\Xset_1,\Xset_2,P_{Y|X_1,X_2},\mathcal{Y})$ consists of finite input alphabets $\Xset_1$, $\Xset_2$, a finite output alphabet $\mathcal{Y}$, and a collection of conditional probability mass functions 
$P_{Y|X_1,X_2}(\cdot|x_1,x_2)$ on $\Yset$, for every $(x_1,x_2)\in\Xset_1 \times \Xset_2$.
Having assumed that the channel is memoryless,  if the inputs $x_1^n=(x_{1}[i])_{i=1}^n$ and $x_2^n=(x_{2}[i])_{i=1}^n$ are sent through $n$ channel uses, then the output is distributed according to 
$P_{Y|X_1,X_2}^n(y^n|x_1^n,x_2^n)=\prod_{i=1}^n P_{Y|X_1,X_2}\big(y[i] \,\big| x_1[i],x_2[i] \big)$.
 The transmitters and the receiver are often called Alice 1, Alice 2, and Bob.

{
\setlength{\tabcolsep}{18pt}
\renewcommand{\arraystretch}{2.75}
\begin{table*}
\caption{Capacity Notations}
\label{Table:C_notation}
\begin{center}
\begin{tabular}{ | m{12em} | m{4cm}|  } 
  \hline
  Notation & Assisting Resources 
  \\   
  \hline
  $\mathcal{C}_{\text{CRT}}(P_{Y|X_1,X_2})$ & Classical common randomness \mbox{between} the transmitters  \\   
  \hline
  $\mathcal{C}_{\text{ET}}(P_{Y|X_1,X_2})$ & Pre-shared entanglement resources between the transmitters  
  \\   
  \hline
  $\mathcal{C}_{\text{ET-C}}(P_{Y|X_1,X_2},C_{12},C_{21})$ & Pre-shared entanglement resources and \emph{classical} conferencing between the transmitters 
    \\   
  \hline
  $\mathcal{C}_{\text{ET-Q}}(P_{Y|X_1,X_2},Q_{12},Q_{21})$ & Pre-shared entanglement resources and \emph{quantum} conferencing between the transmitters   \\ 
  \hline
\end{tabular}
\end{center}
\end{table*}
}

\subsection{Coding with Common Randomness}
\label{sec:Coding_CR}

We begin with the definition of a code for the classical MAC $P_{Y|X_1,X_2}$ with classical CR between the transmitters. We denote the random element that is shared between them  by 
$z$.
\begin{definition} %
\label{def:Clcapacity_CRT}
A $(2^{nR_1},2^{nR_2},n)$   code for the classical MAC $P_{Y|X_1,X_2}$ with
classical CR between the transmitters  consists of the following: 
\begin{itemize}
\item
a  key set $\mathcal{Z}$ and a probability distribution $p_Z$ on $\mathcal{Z}$;
\item 
two message sets  $[1:2^{nR_1}]$ and $ [1:2^{nR_2}]$, assuming $2^{nR_k}$ is an integer;
\item
a pair of encoding functions, $\mathsf{e}_k:[1:2^{nR_k}]\times\mathcal{Z}\to \mathcal{X}_k^n$ for $k\in \{1,2\}$; and
\item
a decoding function   $g:\Yset^n\to [1:2^{nR_1}]\times [1:2^{nR_2}] $.
\end{itemize}
We denote the code by $\mathscr{C}=(P_Z,\mathsf{e}_1,\mathsf{e}_2,g)$.
\end{definition}

\begin{figure*}[tb]
\includegraphics[scale=0.8,trim={-.75cm 10.5cm 2cm 9cm},clip]{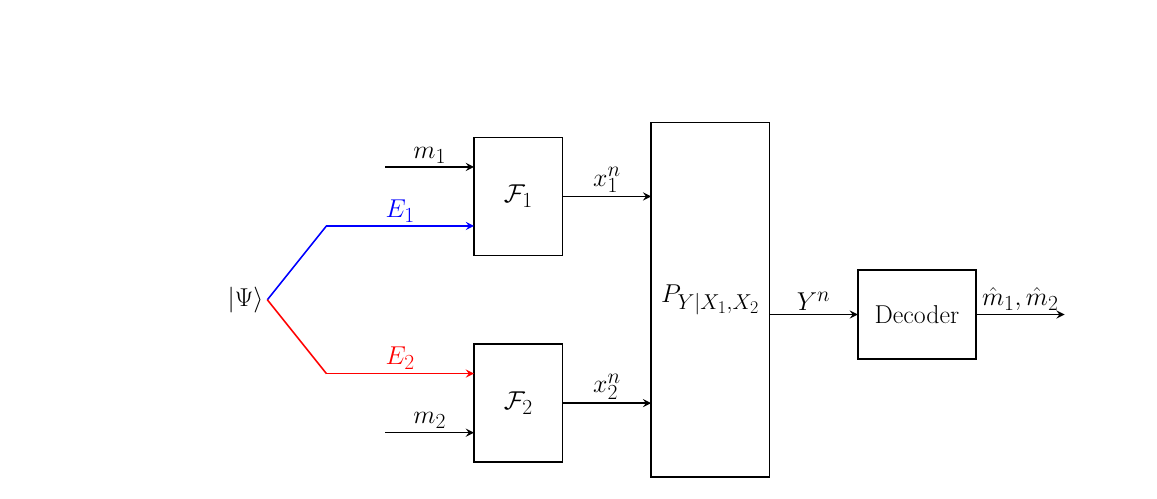} %

\caption{The classical multiple-access channel $P_{Y|X_1,X_2}$ with pre-shared entanglement resources between the transmitters. The entanglement resources (quantum systems) of Transmitter 1 and Transmitter 2 are marked in red and blue, respectively.
}
\label{fig:MentangledTx}
\end{figure*}

The communication scheme is depicted in Figure~\ref{fig:MentangledTx}.  
The transmitters share the classical key  $z\sim p_Z$. 
Alice $k$ chooses a message $m_k$ from the message set, %
$[1:2^{nR_k}]$, for $k=1,2$.
 To send the message $m_1\in [1:2^{nR_1}]$,
Alice 1 transmits $x_1^n=\mathsf{e}_1(m_1,z)$  through $n$ uses of the classical MAC $P_{Y|X_1,X_2}$. %
Similarly, Alice 2  transmits $x_2^n=\mathsf{e}_2(m_2,z)$. 
Bob receives the channel output $y^n$, and estimates the message pair as $(\htm_1,\htm_2)=g(y^n)$.

The conditional probability of error of the code $\mathscr{C}=(P_Z,\mathsf{e}_1,\mathsf{e}_2,g)$ is
\begin{align}
&P_{e}^{(n)}(\mathscr{C}|m_1,m_2)=\sum_{z\in\mathcal{Z}} p_Z(z)  %
\sum_{y^n: g(y^n)\neq (m_1,m_2)}  P_{Y|X_1,X_2}^n(y^n|\mathsf{e}_1(m_1,z),\mathsf{e}_2(m_2,z)) %
\end{align}
for $(m_1,m_2)\in [1:2^{nR_1}]\times [1:2^{nR_2}]$.
 Hence, the maximal probability of error is 
\begin{align}
&P_{e}^{(n)}(\mathscr{C})\equiv %
\max_{m_1,m_2}    P_{e}^{(n)}(\mathscr{C}|m_1,m_2) \,.
\label{Equation:Message_Max_Error_CR}
\end{align}
A rate pair $(R_1,R_2)$ is called achievable with classical  CR between the transmitters   if for every $\eps>0$ and sufficiently large $n$, there exists a 
$(2^{nR_1},2^{nR_2},n)$ code such that the maximal probability of error is bounded by $P_{e}^{(n)}(\mathscr{C})\leq\eps $. 
 The capacity region $\opC_{\text{CRT}}(P_{Y|X_1,X_2})$  
is defined as the closure of the set of achievable pairs $(R_1,R_2)$, where the subscript `CRT' indicates CR resources between the transmitters. %
See  Table~\ref{Table:C_notation}.

\begin{remark}
\label{Remark:Classical_Maximal_Error_CRT}
In general, achievable communication rates may also depend on the error criterion. In particular, we say that
 a rate pair $(R_1,R_2)$ is achievable with 
a \emph{message-average} error criterion 
 if for every $\eps>0$ and sufficiently large $n$, there exists a 
$(2^{nR_1},2^{nR_2},n)$ code such that
$%
\overline{P}_{e}^{(n)}(\mathscr{C})\leq\eps $, where $\overline{P}_{e}^{(n)}(\mathscr{C})$ denotes the message-average error probability:
\begin{align}
&\overline{P}_{e}^{(n)}(\mathscr{C})\equiv %
\frac{1}{2^{n(R_1+R_2)}}\sum_{m_1=1}^{2^{nR_1}}\sum_{m_2=1}^{2^{nR_2}}    P_{e}^{(n)}(\mathscr{C}|m_1,m_2) \,.
\label{Equation:Message_Avg_Error}
\end{align}
The capacity region with a message-average error criterion is defined accordingly.
Without assisting resources, %
the capacity region with a message-average error criterion can be strictly larger than with a maximal error criterion \cite{Dueck:78p} \cite[Sec. 2.2]{Cai:14p}.
For a general MAC without assisting resources, the capacity region under a maximal error criterion is unknown. 
Nevertheless, Cai \cite{Cai:14p} has shown that given classical CR between the transmitters, this is not the case. 
That is, the capacity region with classical CR between the transmitters remains the same, whether we consider a message-average or a maximal error criterion.
\end{remark}

\subsection{Related Work}
The capacity theorem for  classically correlated transmitters is given below.
\begin{theorem}[see also \cite{Cai:14p}]
\label{theo:crtMAC_C}
The capacity region of a classical MAC $P_{Y|X_1,X_2}$ with classical CR between the transmitters is given by
\begin{align}
\mathcal{C}_{\text{CRT}}(P_{Y|X_1,X_2})=
\bigcup_{ p_{V_0} p_{X_1|V_0} p_{X_2|V_0} }
\left\{ \begin{array}{rl}
  (R_1,R_2) \,:\;
	R_1 &\leq I(X_1;Y|X_2 V_0 )  \\
  R_2 &\leq I(X_2;Y|X_1 V_0 ) \\
	R_1+R_2 &\leq I(X_1 X_2;Y|V_0)
	\end{array}
\right\}
\label{Equation:Capacity_CRT}
\end{align}
where the union on the right-hand side 
is over the set of all classical auxiliary variables 
$V_0\sim p_{V_0}$ and all classical channels
$p_{X_k|V_0}$ for $k\in\{1,2\}$. 
\end{theorem}

We note that given such an auxiliary variable $V_0$ and encoding channels $p_{X_k|V_0}$, the variables $V_0\Cbar (X_1,X_2)\Cbar Y$ form a Markov chain: 
\begin{align}
p_{V_0,X_1, X_2, Y}(v_0,x_1,x_2,y)=%
 p_{V_0}(v_0)\cdot p_{X_1|V_0}(x_1|v_0)p_{X_2|V_0}(x_2|v_0)\cdot P_{Y|X_1,X_2}(y|x_1,x_2) \,.
\label{eq:inRsc_Distribution_CR}
\end{align}

The auxiliary variable $V_0$ in the characterization above can thus be interpreted as the CR key that is shared between the transmitters.
From a mathematical perspective,
the union over $p_{V_0}$ can be viewed as a convex hull operation, hence the  capacity region is convex \cite{CoverThomas:06b}.

\begin{remark}
\label{Remark:Time_Sharing}
Consider a message-average error criterion, instead of the maximal error criterion in Definition~\ref{def:Clcapacity_CRT}
(see Remark~\ref{Remark:Classical_Maximal_Error_CRT}).
Then, the convex hull above %
can also be achieved without CR through operational time sharing \cite[Sec. 15.3.3]{CoverThomas:06b}. Thereby, the capacity region with uncorrelated transmitters, i.e., without resources, is \emph{the same} as 
$\mathcal{C}_{\text{CRT}}(P_{Y|X_1,X_2})$.
That is, classical correlation cannot increase the achievable rates, while entanglement can.
The advantage of entanglement will be demonstrated in Subsection~\ref{Subsection:Magic_Example}, based on the examples by Leditzky et al.  \cite{LeditzkyAlhejjiLevinSmith:20p}.
\end{remark}

\renewcommand{\arraystretch}{1}
 \section{Definitions
 }
 \label{Section:Preliminaries}

In the previous section, we have reviewed previously established results on the  MAC with classical CR between the transmitters.
Here, we describe three settings that involve 
entangled transmitters.
  First, we define coding with   entanglement resources between the transmitters.
 Then, we add either classical or quantum conferencing links between the transmitters, in addition to the entanglement resources. 
 The corresponding definitions are given in   in Subsections~\ref{sec:Coding}, \ref{sec:Coding_Conferencing}, and \ref{sec:Coding_Conferencing_Q} below, respectively.
 The notations of the capacity regions are summarized in 
 Table~\ref{Table:C_notation}.

\subsection{Basic Definitions%
}
\label{subsec:notation}

The classical MAC $P_{Y|X_1,X_2}$ is defined as in Subsection~\ref{Subsection:Classical_Notation}.
We use the same notation for classical systems as well.
In addition, we introduce the following notation for quantum systems.
%
%
%
The state of a quantum system $A$ is a density operator $\rho$ on the Hilbert space $\Hset_A$.
 The set of all density operators acting on $\Hset_A$ is denoted by $\mathscr{D}(\Hset_A)$. %
Measurement distributions are
described in terms of a positive operator-valued measure (POVM), i.e., a set of positive semi-denfinite operators  $\{ K_x \}_{x\in\Xset}$
such that $\sum_x K_x=\identity$.
According to the Born rule, if the system is in state $\rho$, then the probability to measure $x$ is  $p_X(x)=\trace(K_x \rho)$.
A quantum channel $\Pset_{A\to B}$ is a completely-positive trace-preserving (cptp) linear map from $\mathscr{D}(\Hset_A)$ to $\mathscr{D}(\Hset_B)$.
A measurement channel is a quantum-classical channel $\mathcal{K}_{A\to X}$, mapping 
$\rho\mapsto\mathcal{K}_{A\to X}(\rho)=\sum_{x\in\mathcal{X}} \trace(K_x \rho)\ketbra{x}$, where $\{ K_x \}_{x\in\Xset}$ forms a POVM
\cite[Def. 4.6.7]{Wilde:17b}.

%
%

\subsection{Coding with Entangled Transmitters}
\label{sec:Coding}
We consider coding for the classical MAC $P_{Y|X_1,X_2}$ with entanglement resources between the transmitters.
This corresponds to the second row in  Table~\ref{Table:C_notation}.
We denote the entanglement resources of Transmitter 1 and Transmitter 2 by $E_1$ and $E_2$, respectively. Let 
$\Hset_{E_1 E_2}\equiv \Hset_{E_1}\otimes \Hset_{E_2}$ denote the corresponding Hilbert space.

\begin{definition} %
\label{def:ClcapacityE}
A $(2^{nR_1},2^{nR_2},n)$   code for the classical MAC $P_{Y|X_1,X_2}$ with
entangled transmitters  consists of the following: 
\begin{itemize}
\item
a bipartite state $\Psi_{E_1 E_2}\in\mathscr{D}(\Hset_{E_1 E_2})$ that is shared between the transmitters.
\item 
two message sets  $[1:2^{nR_1}]$ and $ [1:2^{nR_2}]$, assuming $2^{nR_k}$ is an integer;
\item
two collections of encoding POVMs $\Fset_1^{(m_1)}=\{ F^{(m_1)}_{x_1^n} \}_{x_1^n\in\mathcal{X}_1^n}$ and 
$\Fset_2^{(m_2)}=\{ F^{(m_2)}_{x_2^n} \}_{x_2^n\in\Xset_2^n}$ on $E_1$ and $E_2$, respectively, one pair of POVMs for each message pair $(m_1,m_2)\in [1:2^{nR_1}]\times [1:2^{nR_2}]$; and
\item
a decoding function   $g:\Yset^n\to [1:2^{nR_1}]\times [1:2^{nR_2}] $.
\end{itemize}
We denote the code by $\mathscr{C}=(\Psi,\Fset_1,\Fset_2,g)$.
\end{definition}

The communication scheme is depicted in Figure~\ref{fig:MentangledTx}.  
The transmitters share the entangled pair  $E_1 E_2$. 
Alice $k$ chooses a message $m_k$ from the message set, %
$[1:2^{nR_k}]$, for $k=1,2$.
 To send the message $m_1\in [1:2^{nR_1}]$,
Alice 1 applies the encoding measurement $\Fset_1^{(m_1)}$ to her share of the entanglement resource, $E_1$, and obtains a measurement outcome $x_1^n\in\mathcal{X}_1^n$.
She sends $x_1^n$ through $n$ uses of the classical MAC $P_{Y|X_1,X_2}$. %
Similarly, Alice 2 measures the system $E_2$ using the measurement $\Fset_2^{(m_2)}$, and transmits the outcome $x_2^n$. 
The joint input distribution is thus
\begin{align}%
f(x_1^n,x_2^n|m_1,m_2)
= \trace \left[  \left(F^{(m_1)}_{x_1^n}\otimes F^{(m_2)}_{x_2^n} \right) \Psi_{E_1 E_2} \right] \,.
\end{align}
Bob receives the channel output $y^n$, and estimates the message pair as $(\htm_1,\htm_2)=g(y^n)$.

The conditional probability of error of the code $\mathscr{C}=(\Psi,\Fset_1,\Fset_2,g)$ is
\begin{align}
&P_{e}^{(n)}(\mathscr{C}|m_1,m_2)= %
\sum_{y^n: g(y^n)\neq (m_1,m_2)} \left[  \sum_{(x_1^n,x_2^n)\in \Xset_1^n\times \Xset_2^n} f(x_1^n,x_2^n|m_1,m_2) P_{Y|X_1,X_2}^n(y^n|x_1^n,x_2^n) \right] %
\end{align}
for $(m_1,m_2)\in [1:2^{nR_1}]\times [1:2^{nR_2}]$.
 Hence, the maximal probability of error is 
\begin{align}
&P_{e}^{(n)}(\mathscr{C})\equiv %
\max_{m_1,m_2}    P_{e}^{(n)}(\mathscr{C}|m_1,m_2) \,.
\label{Equation:Message_Max_Error}
\end{align}
A rate pair $(R_1,R_2)$ is called achievable with  entangled transmitters   if for every $\eps>0$ and sufficiently large $n$, there exists a 
$(2^{nR_1},2^{nR_2},n)$ code such that the maximal probability of error is bounded by $P_{e}^{(n)}(\mathscr{C})\leq\eps $. 
 The capacity region $\opC_{\text{ET}}(P_{Y|X_1,X_2})$ of the classical MAC with entangled transmitters
is defined as the closure of the set of achievable pairs $(R_1,R_2)$, where the subscript `ET' indicates entanglement resources between the transmitters. %

\begin{remark}
\label{Remark:Classical_Maximal_Error}
As pointed out in Remark~\ref{Remark:Classical_Maximal_Error_CRT}, achievable communication rates may also depend on the error criterion, and
the capacity region under a maximal error criterion is unknown
for a general MAC without assisting resources.
We also mentioned in Remark~\ref{Remark:Time_Sharing} that Cai \cite{Cai:14p} has shown that this is not the case when the transmitters share classical correlation.
Similarly, the capacity region with entangled transmitters remains the same, whether we consider a message-average or a maximal error criterion.
For completeness, we provide the proof for this property in the appendix.
%
\end{remark}

\begin{figure*}[tb]
\includegraphics[scale=0.8,trim={-.75cm 10.5cm 2cm 9cm},clip]{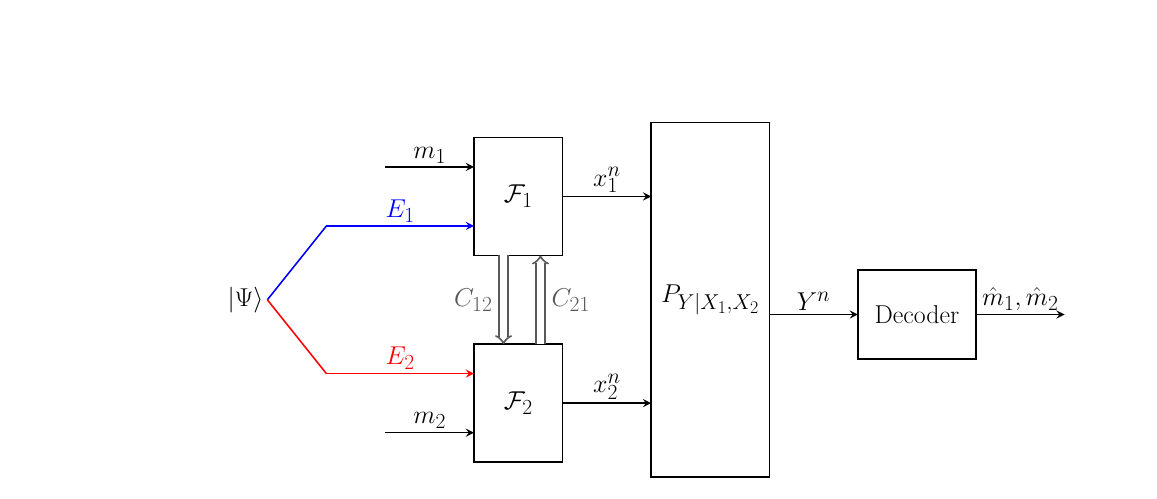} %

\caption{The classical multiple-access channel $P_{Y|X_1,X_2}$ with pre-shared entanglement resources and classical conferencing between the transmitters. The entanglement resources  of Transmitter 1 and Transmitter 2 are marked in blue and red, respectively. The conferencing links are indicated by double-line arrows.
}
\label{fig:MentangledTx_Conferencing}
\end{figure*}

\subsection{Classical Conferencing Cooperation}
\label{sec:Coding_Conferencing}
Now, we  consider coding for the classical MAC $P_{Y|X_1,X_2}$ with both entanglement resources and classical conferencing between the transmitters. 
The capacity notation is as in the third row of  Table~\ref{Table:C_notation}.
As before, the transmitters share entanglement resources on a product Hilbert space
$\Hset_{E_1 E_2}\equiv \Hset_{E_1}\otimes \Hset_{E_2}$. %
Now, suppose that given the messages, the transmitters may communicate with each other over noise-free classical links at a limited rate. The link from Transmitter 1 to Transmitter 2 has a classical capacity of $C_{12}$, and from Transmitter 2 to Transmitter 1 has a classical capacity of 
$C_{21}$, where $C_{12}$ and $C_{21}$ are fixed. In this case, we say that the MAC has a classical $(C_{12},C_{21})$-permissible  conference between the transmitters.

\begin{definition} %
\label{def:ClcapacityE_Conferencing}
A $(2^{nR_1},2^{nR_2},n)$   code for the classical MAC $P_{Y|X_1,X_2}$ with
entangled transmitters
and classical  conferencing 
consists of the following: 
\begin{itemize}
\item
a bipartite state $\Psi_{E_1 E_2}\in\mathscr{D}(\Hset_{E_1 E_2})$ that is shared between the transmitters.
\item 
two message sets  $[1:2^{nR_1}]$ and $ [1:2^{nR_2}]$, %
\item 
two sets of $T$ classical conferencing functions
\begin{align}
s_{1\to 2}^{(t)}&:[1:2^{nR_1}]
\times 
\left( \Omega_{2\to 1}^{(1)}\times \cdots\times
 \Omega_{2\to 1}^{(t-1)} \right)
 \to
\Omega_{1\to 2}^{(t)}
 \,,
\\
s_{2\to 1}^{(t)}&: [1:2^{nR_2}]
\times 
\left(\Omega_{1\to 2}^{(1)}\times \cdots\times
\Omega_{1\to 2}^{(t-1)} \right)
\to
\Omega_{2\to 1}^{(t)}
\end{align}
for $t\in [1:T]$, where $\Omega_{1\to 2}^{(t)}$ and $\Omega_{2\to 1}^{(t)}$ are finite conferencing alphabets, and $\Omega_{1\to 2}^{(0)}=\Omega_{2\to 1}^{(0)}=\emptyset$.

\item
two collections of encoding POVMs $\Fset_1^{(m_1,\omega_{2\to 1}^T)}=\{ F^{(m_1,\omega_{2\to 1}^T)}_{x_1^n} \}_{x_1^n\in\mathcal{X}_1^n}$ and 
$\Fset_2^{(m_2,\omega_{1\to 2}^T)}=\{ F^{(m_2,\omega_{1\to 2}^T)}_{x_2^n} \}_{x_2^n\in\Xset_2^n}$ on $E_1$ and $E_2$, respectively; and
\item
a decoding function   $g:\Yset^n\to [1:2^{nR_1}]\times [1:2^{nR_2}] $.
\end{itemize}
We denote the code by $\mathscr{C}=(\Psi,s_{1\to 2},s_{2\to 1},\Fset_1,\Fset_2,g)$.
\end{definition}

The communication scheme is depicted in Figure~\ref{fig:MentangledTx_Conferencing}.  
The transmitters share the entangled pair  $E_1 E_2$. 
Alice $k$ chooses a message $m_k$ from the message set, %
$[1:2^{nR_k}]$, for $k=1,2$.
Then, Alice 1 and Alice 2 perform a classical conferencing protocol as follows.
In round $t=1$,
Alice 1 sends a conferencing message 
$
\omega_{1\to 2}[1]=s_{1\to 2}^{(1)}(m_1)
$
to Alice 2. At the same time, Alice 2 sends a conferencing message 
$
\omega_{2\to 1}[1]=s_{2\to 1}^{(1)}(m_2)
$
to Alice 1. Next,
in round $t\in [2:T]$, Alice 1 sends a conferencing message 
$
\omega_{1\to 2}[t]=s_{1\to 2}^{(t)}\left(m_1,\omega_{2\to 1}[1],\ldots,\omega_{2\to 1}[t-1]\right)
$
to Alice 2, meanwhile Alice 2 sends a conferencing message 
$
\omega_{2\to 1}[t]=s_{2\to 1}^{(t)}\left(m_2,\omega_{1\to 2}[1],\ldots,\omega_{1\to 2}[t-1]\right)
$
to Alice 1.

 To send the message $m_1\in [1:2^{nR_1}]$ to the receiver,
Alice 1 applies the encoding measurement $\Fset_1^{(m_1,\omega_{2\to 1}^T)}$ to her share of the entanglement resource, $E_1$, and obtains a measurement outcome $x_1^n\in\mathcal{X}_1^n$.
She sends $x_1^n$ through $n$ uses of the classical MAC $P_{Y|X_1,X_2}$. %
Similarly, Alice 2 measures the system $E_2$ using the measurement $\Fset_2^{(m_2,\omega_{1\to 2}^T)}$, and transmits the outcome $x_2^n$. 
The joint input distribution given the information messages and the conferencing message is then
\begin{align}%
f(x_1^n,x_2^n|m_1,m_2,\omega_{1\to 2}^T,\omega_{2\to 1}^T)
= \trace \left[  \left(F^{(m_1,\omega_{2\to 1}^T)}_{x_1^n}\otimes F^{(m_2,\omega_{1\to 2}^T)}_{x_2^n} \right) \Psi_{E_1 E_2} \right] \,.
\end{align}%
Note that the measurement of Alice 1 depends on the received conferencing messages $\omega_{2\to 1}^T$, which in turn, depends on the message $m_2$ of Alice 2.  Similarly, Alice 2's measurement depends on $m_1$ through message $\omega_{1\to 2}^T$. Therefore, the correlation between the transmitters is aided by both the shared entanglement and  classical conferencing.
Bob receives the channel output $y^n$, and estimates the message pair as $(\htm_1,\htm_2)=g(y^n)$.

Let $C_{12}$ and $C_{21}$ be fixed non-negative constants.
The code is  $(C_{12},C_{21})$-permissible if
\begin{align}
\abs{\overline{\Omega}^{T}_{1\to 2}}%
\leq  2^{nC_{12}}
\text{ , and }\;
\abs{\overline{\Omega}^{T}_{2\to 1}}\leq  2^{nC_{21}}
\label{eq:Permissible}
\end{align}
where $\overline{\Omega}^{T}_{k\to j}=%
\Omega^{(1)}_{k\to j}\times\cdots\times \Omega^{(T)}_{k\to j}
$ is the overall conference message set from Alice $k$ to Alice $j$. %
We denote the  $T$-sequences of conferencing messages, sent from Alice 1 to Alice 2, and from Alice 2 to Alice 1, by 
$\omega_{1\to 2}^T\equiv \overline{s}_{1\to 2}^T(m_1,m_2)$ and $\omega_{2\to 1}^T\equiv \overline{s}_{2\to 1}^T(m_1,m_2)$, respectively, where 
\begin{align}
\overline{s}_{1\to 2}^T(m_1,m_2)&=\left[
s_{1\to 2}^{(t)}\left(m_1,s_{2\to 1}^{(t-1)}\right)
\right]_{t=1}^T \,,
\intertext{and }
\overline{s}_{2\to 1}^T(m_1,m_2)&=\left[
s_{2\to 1}^{(t)}\left(m_2,s_{1\to 2}^{(t-1)}\right)
\right]_{t=1}^T \,.
\end{align}

The conditional probability of error of the code $\mathscr{C}=(\Psi,s_{1\to 2},s_{2\to 1},\Fset_1,\Fset_2,g)$ is
\begin{multline}
P_{e}^{(n)}(\mathscr{C}|m_1,m_2)= %
\sum_{y^n: g(y^n)\neq (m_1,m_2)} 
  \sum_{(x_1^n,x_2^n)\in \Xset_1^n\times \Xset_2^n}   
  f(x_1^n,x_2^n|m_1,m_2,\overline{s}_{1\to 2}^T(m_1,m_2),\overline{s}_{2\to 1}^T(m_1,m_2))
\\ \cdot
P_{Y|X_1,X_2}^n(y^n|x_1^n,x_2^n) %
\end{multline}
for $(m_1,m_2)\in [1:2^{nR_1}]\times [1:2^{nR_2}]$.
The maximal probability of error, $P_{e}^{(n)}(\mathscr{C})$, is defined as before (see (\ref{Equation:Message_Max_Error})).

A rate pair $(R_1,R_2)$ is called achievable with  entangled  transmitters and classical conferencing  if for every $\eps>0$ and sufficiently large $n$, there exists a 
$(C_{12},C_{21})$-permissible code such that 
$P_{e}^{(n)}(\mathscr{C})\leq\eps $.
 The capacity region $\opC_{\text{ET-C}}(P_{Y|X_1,X_2},C_{12},C_{21})$ %
is defined as in Section~\ref{sec:Coding}, %
where the subscript  indicates entanglement resources  and classical conferencing between the transmitters. %

\begin{figure*}[tb]
\includegraphics[scale=0.8,trim={-.75cm 10.5cm 2cm 9cm},clip]{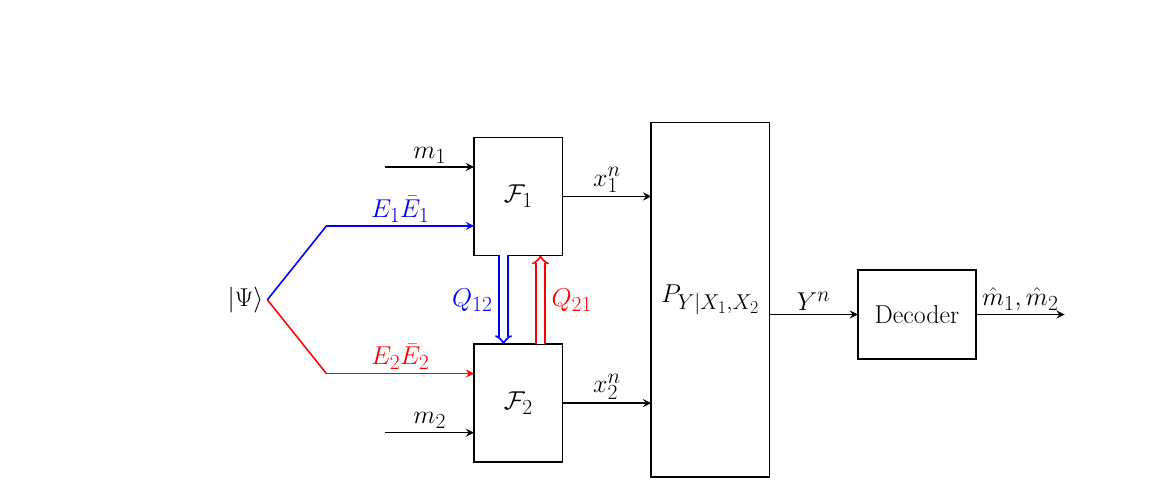} %

\caption{The classical multiple-access channel $P_{Y|X_1,X_2}$ with pre-shared entanglement resources and quantum conferencing between the transmitters. The quantum systems  of Transmitter 1 and Transmitter 2 are marked in blue and red, respectively. 
}
\label{fig:MentangledTx_Conferencing_Q}
\end{figure*}

\subsection{Quantum Conferencing with Entanglement}
\label{sec:Coding_Conferencing_Q}
Next, we consider coding for the classical MAC $P_{Y|X_1,X_2}$ with quantum conferencing, while entanglement resources are available at both the conferencing stage and the transmission. 
The capacity notation is as in the last row of
 Table~\ref{Table:C_notation}.
Here, we denote the entanglement resources that are being used during the conferencing protocol by  $\bar{E}_1$ and $\bar{E}_2$, and during the transmission by $E_1 $ and $E_2$, corresponding to 
 Transmitter 1 and Transmitter 2, respectively.  
Hence, the transmitters share entanglement resources on a product Hilbert space
$\Hset_{E_1 \bar{E}_1 E_2  \bar{E}_2}$. %
Now, suppose that given the messages, the transmitters may communicate with each other over noise-free quantum links at a limited rate with entanglement assistance. The link from Transmitter 1 to Transmitter 2 has a quantum capacity of $Q_{12}$, and from Transmitter 2 to Transmitter 1 has a quantum capacity of 
$Q_{21}$, where $Q_{12}$ and $Q_{21}$ are fixed. In this case, we say that the MAC has a $(Q_{12},Q_{21})$-permissible conference between the transmitters.

To simplify the model, we consider a single round of quantum conferencing.

\begin{definition} %
\label{def:ClcapacityE_Conferencing_Q}
A $(2^{nR_1},2^{nR_2},n)$   code for the classical MAC $P_{Y|X_1,X_2}$ with
entanglement-assisted   conferencing transmitters
consists of the following: 
\begin{itemize}
\item
a bipartite state $\Psi_{E_1 E_2 \bar{E}_1 \bar{E}_2}\in\mathscr{D}(\Hset_{E_1 E_2 \bar{E}_1 \bar{E}_2})$ that is shared between the transmitters.
\item 
two message sets  $[1:2^{nR_1}]$ and $ [1:2^{nR_2}]$, %
\item 
two conferencing quantum channels $\mathcal{S}_{1\to 2}^{(m_1)}:\mathscr{D}(\mathcal{H}_{\bar{E}_1})\to \mathscr{D}(\mathcal{H}_{D_{1\to 2}})$ and
$\mathcal{S}_{2\to 1}^{(m_2)}:\mathscr{D}(\mathcal{H}_{\bar{E}_2})\to \mathscr{D}(\mathcal{H}_{D_{2\to 1}})$,

\item
two collections of encoding POVMs $\Fset_1^{(m_1)}=\{ F^{(m_1)}_{x_1^n} \}_{x_1^n\in\mathcal{X}_1^n}$ and 
$\Fset_2^{(m_2)}=\{ F^{(m_2)}_{x_2^n} \}_{x_2^n\in\Xset_2^n}$ on $E_1 D_{2\to 1}$ and $E_2 D_{1\to 2}$, respectively; and
\item
a decoding function   $g:\Yset^n\to [1:2^{nR_1}]\times [1:2^{nR_2}] $.
\end{itemize}
We denote the code by $\mathscr{C}=(\Psi,\mathcal{S}_{1\to 2},\mathcal{S}_{2\to 1},\Fset_1,\Fset_2,g)$.
\end{definition}

The communication scheme is depicted in Figure~\ref{fig:MentangledTx_Conferencing_Q}.  
The transmitters share the entangled pair of 
$E_1 \bar{E}_1$ and $ E_2 \bar{E}_2$. 
Alice $k$ chooses a message $m_k$ from the message set, %
$[1:2^{nR_k}]$, for $k=1,2$.
Then, Alice 1 and Alice 2 perform a quantum  conferencing protocol as follows.
Alice 1 applies the quantum conferencing channel $\mathcal{S}_{1\to 2}^{(m_1)}$ to her entangled resource $\bar{E}_1$, and sends the output system $D_{1\to 2}$ to Alice 2 through her quantum conferencing link. 
At the same time, Alice 2 
applies  $\mathcal{S}_{2\to 1}^{(m_2)}$ to  $\bar{E}_2$, and sends the output $D_{2\to 1}$ to Alice 1. At the end of the quantum conferencing protocol, we have a joint state
\begin{align}
\sigma_{E_1 E_2 D_{1\to 2} D_{2\to 1}}=
(\mathrm{id}_{E_1 E_2}\otimes \mathcal{S}_{1\to 2}^{(m_1)}\otimes \mathcal{S}_{2\to 1}^{(m_2)})(\Psi_{E_1 E_2 \bar{E}_1 \bar{E}_2})
\end{align}
Notice that the conferencing protocol exploits the entanglement resources $\bar{E}_1$ and $\bar{E}_2$ to encode $D_{1\to 2}$ and $D_{2\to 1}$, respectively. The remaining resources, $E_1$ and $E_2$, will be used in the next stage.

 To send the message $m_1\in [1:2^{nR_1}]$ to the receiver,
Alice 1 performs a joint measurement $\Fset_1^{(m_1)}$ on the entangled resource, $E_1$, together with the quantum conference system $D_{2\to 1}$, and obtains a measurement outcome $x_1^n\in\mathcal{X}_1^n$.
She sends $x_1^n$ through $n$ uses of the classical MAC $P_{Y|X_1,X_2}$. %
Similarly, Alice 2 performs a measurement $\Fset_2^{(m_2)}$ on  $E_2$ and $D_{1\to 2}$, and transmits the outcome $x_2^n$. 
The joint input distribution is then
\begin{align}%
f(x_1^n,x_2^n|m_1,m_2)
= \trace \left[  \left(F^{(m_1)}_{x_1^n} \otimes F^{(m_2)}_{x_2^n} \right) \sigma_{E_1 E_2 D_{1\to 2} D_{2\to 1}} \right] \,.
\end{align}%
Note that the measurement of Alice 1 includes the conferencing system $D_{2\to 1}$, which in turn, is correlated with the message $m_2$ of Alice 2.  Similarly, Alice 2's measurement includes a conferencing system that is correlated with $m_1$. Therefore, the correlation between the transmitters is aided by both the shared entanglement and  quantum conferencing.
Bob decodes as usual and estimates  $(\htm_1,\htm_2)=g(y^n)$.

Let $Q_{12}$ and $Q_{21}$ be fixed non-negative constants.
The code is  $(Q_{12},Q_{21})$-permissible if
\begin{align}
\mathrm{dim}(\mathcal{H}_{D_{1\to 2}})
\leq  2^{nQ_{12}}
\text{ , and }\;
\mathrm{dim}(\mathcal{H}_{D_{2\to 1}})\leq  2^{nQ_{21}}
\,.
\label{eq:Permissible_Q}
\end{align}
The maximal probability of error, $P_{e}^{(n)}(\mathscr{C})$, is defined as before (see (\ref{Equation:Message_Max_Error})).

A rate pair $(R_1,R_2)$ is called achievable with  entangled  conferencing transmitters  if for every $\eps>0$ and sufficiently large $n$, there exists a 
$(Q_{12},Q_{21})$-permissible code such that 
$P_{e}^{(n)}(\mathscr{C})\leq\eps $.
 The capacity region $\opC_{\text{ET-Q}}(P_{Y|X_1,X_2},Q_{12},Q_{21})$ of the classical MAC with entangled  conferencing transmitters
is defined as the closure of the set of achievable pairs $(R_1,R_2)$, where the subscript `ET-Q' indicates entanglement resources  and quantum conferencing between the transmitters. %

Before we move on to our results, we summarize the definitions given so far. 
Overall, we have defined four settings of the MAC with assisting resources between the transmitters. 
Definition~\ref{def:Clcapacity_CRT}, in the previous section, describes coding with classical CR.
Here, in this section, we have considered coding with pre-shared entanglement resources between the transmitters.
Definition~\ref{def:ClcapacityE} above considers coding with pre-shared entanglement resources between the transmitters.
In Definition~\ref{def:ClcapacityE_Conferencing}, we add classical conferencing, where $C_{12}$ and $C_{21}$ are the conferencing rates from
Transmitter~1  to Transmitter~2, and vice versa, respectively. 
Definition~\ref{def:ClcapacityE_Conferencing_Q} above introduces the notion of quantum conferencing between the transmitters, at quantum rates  $Q_{12}$ and $Q_{21}$. The corresponding capacity notation is shown in Table~\ref{Table:C_notation}.

\begin{remark}

 The resources in models that include entanglement resources are \emph{stronger} than classical CR, as entanglement can be used in order to generate classical CR. 
 Furthermore, 
 we have
 \begin{align}
\opC_{\text{CR}}(P_{Y|X_1,X_2})
&\subseteq
\opC_{\text{ET}}(P_{Y|X_1,X_2})
\nonumber\\&
\subseteq
\opC_{\text{ET-C}}(P_{Y|X_1,X_2},C_{12},C_{21})
\Big|_{\substack{C_{12}=r\\C_{21}=r'}}
\nonumber\\&
\subseteq
\opC_{\text{ET-Q}}(P_{Y|X_1,X_2},Q_{12},Q_{21})
\Big|_{\substack{Q_{12}=r \\Q_{21}=r'}}
 \end{align}
 for every $r,r'>0$.
\end{remark}

\begin{remark}
We focus here on two conferencing models. In the discussion section, we will consider other variations.  
See Subsection~\ref{Subsection:Open}. 
\end{remark}

\section{Main Results --- Entangled Transmitters}
\label{Section:Main_Results}

\subsection{Communication With Unlimited Entanglement Resources}
\label{Subsection:Main_Result}
We state our results on the classical MAC $P_{Y|X_1,X_2}$ with entanglement resources between the transmitters.
We present inner and outer bounds on the capacity region, as well as a regularized characterization.

\subsubsection{Inner Bound and Regularized Characterization}
Define the rate region $\mathcal{R}_{\text{ET}}(P_{Y|X_1,X_2})$ as follows,
\begin{align}%
\mathcal{R}_{\text{ET}}(P_{Y|X_1,X_2})&=
\bigcup_{ p_{V_0} p_{V_1|V_0} p_{V_2|V_0} \,,\; \varphi_{A_1 A_2} \,,\; \Lset_1 \otimes \Lset_2 }
\left\{ \begin{array}{rl}
  (R_1,R_2) \,:\;
	R_1 &\leq I(V_1;Y|V_0 V_2)  \\
  R_2 &\leq I(V_2;Y|V_0 V_1) \\
	R_1+R_2 &\leq I(V_1 V_2;Y|V_0)
	\end{array}
\right\} \,,
\label{eq:inRetx_In}
\end{align}%
The union on the right-hand side of (\ref{eq:inRetx_In}) 
is over the set of all
 bipartite states $\varphi_{A_1 A_2}
 \in
\mathscr{D}( \Hset_{A_1}\otimes \Hset_{ A_2})$, classical auxiliary variables $(V_0,V_1,V_2)\sim p_{V_0} p_{V_1|V_0} p_{V_2|V_0}$, and collection of POVMs $\Lset_k(v_0,v_k)=\{L_k(x_k|v_0,v_k)\}_{x_k\in\Xset_k}$, for $v_0\in\Vset_0$, $v_k\in\Vset_k$, $k=1,2$. Note that the classical variables $V_1\Cbar V_0\Cbar V_2$ form a Markov chain.

Given such a state, variables, and POVMs, the joint distribution of $(V_0,V_1,V_2,X_1,X_2,Y)$ is 
\begin{multline}
p_{V_0,V_1,V_2,X_1, X_2, Y}(v_0,v_1,v_2,x_1,x_2,y)=%
 p_{V_0}(v_0) p_{V_1|V_0}(v_1|v_0)p_{V_2|V_0}(v_2|v_0)\\ \cdot\trace\left[ \left( L_1(x_1|v_0,v_1)\otimes L_2(x_2|v_0,v_2) \right)\varphi_{A_1 A_2}  \right] \cdot P_{Y|X_1,X_2}(y|x_1,x_2) \,.
\label{eq:inRsc_Distribution}
\end{multline}
The subscript `ET' stands for entangled transmitters.

Before we state the achievability theorem, we give the following lemma. 
 The property stated below in Lemma~\ref{lemm:CardCl} can simplify the computation  %
 of the region $\mathcal{R}
 _{\text{ET}}(P_{Y|X_1,X_2})$  %
for a given channel.
\begin{lemma}
\label{lemm:CardCl}
The union of the inner bound in (\ref{eq:inRetx_In}) can be attained with %
auxiliary variables $V_0$, $V_1$, $V_2$ 
with $|\Vset_0|\leq 3$, $|\Vset_k|\leq 3(|\Xset_1||\Xset_2|+2)$, $k=1,2$, and 
pure states $\varphi_{A_1 A_2}\equiv \ketbra{\phi_{A_1 A_2}}$.
In other words, the region does not reduce as we limit the union in this manner. 
\end{lemma}
The proof of Lemma~\ref{lemm:CardCl} is based on the Fenchel-Eggleston-Carath\'eodory theorem \cite{Eggleston:66p} and
the perturbation method \cite{GohariAnantharam:12p,GraceGuha:22c}.
The details are given in Section%
~\ref{app:CardCl}. 
\begin{remark}
\label{Remark:Dimension}
In this section, we consider the capacity region given unlimited entanglement resources between the transmitters. 
We note that the dimension of the entangled systems $A_1$ and $A_2$ in the rate formula (\ref{eq:inRetx_In}) is unbounded. As the rate formula
$\mathcal{R}
_{\text{ET}}(P_{Y|X_1,X_2})$ involves a union over the \emph{closure} of the set of finite-dimensional quantum correlations,
we include
 correlations that  are obtained in the limit of 
%
%
$\mathrm{dim}(\mathcal{H}_{A_k})\to\infty$.
This family of correlations is often denoted in the literature by 
$C_{qa}$ \cite{Slofstra:20p}.
As we will discuss in Section~\ref{Section:Discussion}, there are cases where the union cannot be exhausted with finite-dimensional entangled systems \cite{LeditzkyAlhejjiLevinSmith:20p,Slofstra:20p}
(see Example~\ref{Example:Slofstra_Vidick} below).
\end{remark}
Next, we state our achievability result and the regularized characterization.

\begin{theorem}
\label{theo:etMAC_In}
The capacity region of a classical MAC $P_{Y|X_1,X_2}$ with entangled transmitters satisfies
\begin{align}
\mathcal{C}_{\text{ET}}(P_{Y|X_1,X_2})\supseteq \mathcal{R}_{\text{ET}}(P_{Y|X_1,X_2}) \,.
\end{align}
Furthermore, the capacity region is given by 
\begin{align}
\mathcal{C}_{\text{ET}}(P_{Y|X_1,X_2})= \bigcup_{\ell=1}^{\infty} \frac{1}{\ell}\mathcal{R}_{\text{ET}}(P_{Y|X_1,X_2}^{\ell})
\end{align}
where $P_{Y|X_1,X_2}^{\ell}$ is the $\ell$-fold product channel (see notation in Subsection~\ref{Subsection:Classical_Notation}). 
\end{theorem}
The proof of Theorem~\ref{theo:etMAC_In} is given in Section%
~\ref{app:etMAC_In}.
In Subsection~\ref{Subsection:Magic_Example} below, we will review two examples by Leditzky et al. \cite{LeditzkyAlhejjiLevinSmith:20p} that show that the capacity region with entangled transmitters can be strictly larger than without such correlation resources.

\begin{remark}
\label{rem:Uncomput}
In the basic point-to-point communication problem, Bennett et al.  
\cite{BennettShorSmolin:99p,BennettShorSmolin:02p} showed that entanglement assistance between the transmitter and the \emph{receiver} leads to a characterization that is easy to compute.
In other words, in Bennett et al.'s model \cite{BennettShorSmolin:02p}, introducing 
entanglement resources  transforms the capacity
evaluation from an uncomputable task to an optimization
that can be easily performed, numerically
\cite[Remark 5]{PeregDeppeBoche:21p}.
Unfortunately,  this is not the case in the present work. Clearly, our achievable region  has a single-letter form with respect to the channel dependency. However, there is no upper bound on the necessary dimension of the auxiliary systems $A_1$ and $A_2$ in Theorem~\ref{theo:etMAC_In}
(see Remark~\ref{Remark:Dimension}).
In principle, one can  compute achievable rates by choosing an arbitrary dimension, but the best inner bound cannot be  computed with absolute precision in general. 
In the discussion section, we affirm that this is an inherent limitation to our problem
 (see Section~\ref{Section:Discussion}).
 Furthermore, even without entanglement assistance, there exists MACs such that the computation of the capacity region is NP-hard \cite{LeditzkyAlhejjiLevinSmith:20p}.
\end{remark}

\begin{remark}
\label{Remark:Common_Randomness}
In communication, shared entanglement is often thought of as the quantum version of common randomness (CR), which is a random key that is shared between the two parties, see Section~\ref{Section:Related_Work}.
Recall that we denote the capacity region 
with CR-assisted transmitters by $\mathcal{C}_{\text{CRT}}(P_{Y|X_1,X_2})$.
Then, observe that if we choose the measurements in (\ref{eq:inRetx_In}) such that 
$X_k=V_k$ for $k=1,2$, we obtain the capacity region with CR assistance:
\begin{align}%
\mathcal{C}_{\text{ET}}(P_{Y|X_1,X_2})\supseteq
\bigcup_{ p_{V_0} p_{X_1|V_0} p_{X_2|V_0} }
\left\{ \begin{array}{rl}
  (R_1,R_2) \,:\;
	R_1 &\leq I(X_1;Y|X_2 V_0 )  \\
  R_2 &\leq I(X_2;Y|X_1 V_0 ) \\
	R_1+R_2 &\leq I(X_1 X_2;Y|V_0)
	\end{array}
\right\}=\mathcal{C}_{\text{CRT}}(P_{Y|X_1,X_2}) \,.
\label{eq:capacityCR}
\end{align}
where the last equality holds by Theorem~\ref{theo:crtMAC_C} (see \cite{Cai:14p}).
In the examples by Leditzky et al.  \cite{LeditzkyAlhejjiLevinSmith:20p},  there is strict inclusion, i.e.,
$\mathcal{C}_{\text{CRT}}(P_{Y|X_1,X_2})\subsetneqq \mathcal{C}_{\text{ET}}(P_{Y|X_1,X_2})$. 
We will show the examples in Subsection~\ref{Subsection:Magic_Example}.
\end{remark}

\subsubsection{Outer Bound}
Next, we establish an outer bound. Define the rate region
\begin{align}%
&
\mathcal{O}_{\text{ET}}(P_{Y|X_1,X_2})
=
\bigcup_{  p_{V_0 V_1 V_2}  \,,\; \varphi_{A_1 A_2} \,,\; \Lset_1 \otimes \Lset_2 }
\left\{ \begin{array}{rl}
  (R_1,R_2) \,:\;
	R_1 &\leq I(V_1;Y| V_0 V_2)  \\
        R_2 &\leq I(V_2;Y| V_0 V_1) \\
	R_1+R_2 &\leq I(V_1 V_2;Y|V_0)
	\end{array}
\right\} 
\label{eq:inRetx_Out_1}
\end{align}%
The difference between the formula given above  and the inner bound in (\ref{eq:inRetx_In}) is in the joint distribution of the classical auxiliary variables.
Specifically, in (\ref{eq:inRetx_In}), we limit the joint distribution such that
$V_1\Cbar V_0\Cbar V_2$ form a Markov chain.  Here, we remove this restriction and consider any joint distribution. 
The union of the outer bound in (\ref{eq:inRetx_Out_1}) is exhausted by 
 auxiliary variables and pure states as in  Lemma~\ref{lemm:CardCl} following the same considerations. The derivation of those properties is given in Section%
~\ref{app:CardCl_Out}. 
The lack of a dimension bound for the bipartite state extends to the outer bound as well (see Remark~\ref{Remark:Dimension}).

\begin{theorem}
\label{theo:etMAC_Out}
The capacity region of a classical MAC $P_{Y|X_1,X_2}$ with entangled transmitters satisfies
\begin{align}
\mathcal{C}_{\text{ET}}(P_{Y|X_1,X_2})\subseteq \mathcal{O}_{\text{ET}}(P_{Y|X_1,X_2}) \,.
\end{align}
\end{theorem}

\subsection{Rate-Limited Entanglement Resources}
\label{Subsection:Rate_Limit_EA}
Now, we consider a setting where the entanglement resources are limited.
A code with rate-limited entanglement resources
is defined such that 
 the encoders have access to Hilbert spaces which grow as a function of the channel uses in a rate limited fashion.
The precise definition is given below.
\begin{definition} %
\label{def:ClcapacityE_Limited}
A $(2^{nR_1},2^{nR_2},n)$   code for the classical MAC $P_{Y|X_1,X_2}$ with an entanglement rate $\theta_E$ between the
 transmitters  consists of the following: 
\begin{itemize}
\item
a bipartite state $\Psi_{E_1 E_2}$ that is shared between the transmitters, with $\mathrm{dim}(\mathcal{H}_{E_k})\leq 2^{n\theta_E}$ for $k=1,2$.
\item 
two message sets  $[1:2^{nR_1}]$ and $ [1:2^{nR_2}]$, assuming $2^{nR_k}$ is an integer;
\item
a pair of encoding POVMs $\Fset_1^{(m_1)}=\{ F^{(m_1)}_{x_1^n} \}_{x_1^n\in\mathcal{X}_1^n}$ and 
$\Fset_2^{(m_2)}=\{ F^{(m_2)}_{x_2^n} \}_{x_2^n\in\Xset_2^n}$ on $E_1$ and $E_2$, respectively; and
\item
a decoding function   $g:\Yset^n\to [1:2^{nR_1}]\times [1:2^{nR_2}] $.
\end{itemize}
\end{definition}
The communication scheme is as in
Subsection~\ref{sec:Coding}
(see
Figure~\ref{fig:MentangledTx}).  
A rate pair $(R_1,R_2)$ is said to be achievable with  an entanglement rate $\theta_E$    if for every $\eps>0$ and sufficiently large $n$, there exists a 
$(2^{nR_1},2^{nR_2},n)$ code such that the transmitters share entanglement resources at a rate $\theta_E$ and the maximal probability of error is bounded by $P_{e}^{(n)}(\mathscr{C})\leq\eps $. 
 The  capacity region $\opC_{\text{ET}}(P_{Y|X_1,X_2},\theta_E)$ with rate-limited  entanglement between the transmitters
is defined as the set of achievable pairs $(R_1,R_2)$, with an  entanglement rate $\theta_E$ between the transmitters. %

We  note that by definition, the capacity region with unlimited entanglement between the transmitters is $\mathcal{C}_{\text{ET}}(\cdot)\equiv \mathcal{C}_{\text{ET}}(\cdot,+\infty)$.
Therefore, for every entanglement rate
$\theta_E>0$, we have
$\mathcal{C}_{\text{ET}}(\cdot,\theta_E)\subseteq
\mathcal{C}_{\text{ET}}(\cdot)$.

Based on the achievability proof in Section%
~\ref{app:etMAC_In} and Lemma~\ref{lemm:CardCl}, we obtain the following consequence.
\begin{corollary}
\label{coro:etMAC_Limited}
Let $\theta_E>0$ be a given entanglement rate.
A rate pair $(R_1,R_2)$ is achievable with entanglement at rate $\theta_E$ between the transmitters if 
\begin{align}%
\begin{array}{rl}
	R_1 &\leq I(V_1;Y|V_0 V_2)  \\
  R_2 &\leq I(V_2;Y|V_0 V_1) \\
	R_1+R_2 &\leq I(V_1 V_2;Y|V_0)
	\end{array}
\label{eq:inRetx_Limited}
\end{align}%
for a pure state $\ket{\phi_{A_1 A_2}}$
with entropy
\begin{align}
H(A_1)_\phi=H(A_2)_\phi\leq \theta_E \,,
\end{align}
some classical variables $(V_0,V_1,V_2)\sim p_{V_0} p_{V_1|V_0} p_{V_2|V_0}$, 
  and conditional measurement $\Lset_k(v_0,v_k)=\{L_k(x_k|v_0,v_k)\}_{x_k\in\Xset_k}$, for $v_0\in\Vset_0$, $v_k\in\Vset_k$, $k=1,2$. 
 \end{corollary}

\begin{table}
\renewcommand{\arraystretch}{1.3}
\caption{The magic square game: Deterministic strategies}
\label{Table:Magic_Classical}
\centering
\begin{tabular}{| c| c| c| }
\hline
 0 & 0 & 0 \\ 
\hline
 0 & 1 & 1 \\  
\hline
 1 & 0 & ?    \\
\hline
\end{tabular}
\end{table}

\begin{table}
\renewcommand{\arraystretch}{1.3}
\caption{The magic square game: Quantum strategies}
\label{Table:Magic_Quantum}
\centering
\begin{tabular}{| c| c| c| }
\hline
 $\mathsf{X}\otimes\identity$ & $\mathsf{X}\otimes \mathsf{X}$ & $\identity\otimes \mathsf{X}$ \\ 
\hline
 $-\mathsf{X}\otimes \mathsf{Z}$ & $\mathsf{Y}\otimes \mathsf{Y}$ & $-\mathsf{Z}\otimes \mathsf{X}$ \\  
\hline
 $\identity\otimes \mathsf{Z}$ & $\mathsf{Z}\otimes \mathsf{Z}$ & $\mathsf{Z}\otimes \identity$    \\
\hline
\end{tabular}
\end{table}

\begin{figure*}[tb]
\includegraphics[scale=1.85,trim={2cm 21cm 2cm 
4.5cm},clip]{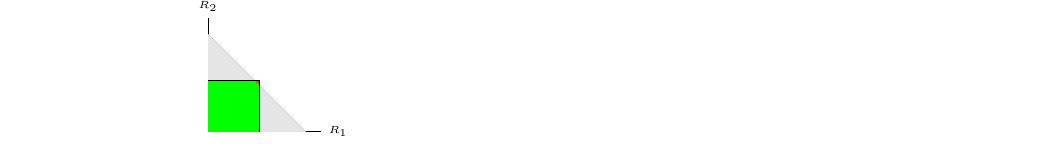} %
\caption{Rate bounds for the magic-square  multiple-access channel $P_{Y|X_1,X_2}$ in Example~\ref{Example:Magic_Square}.
The gray triangle indicates the outer bound for the capacity region $\mathcal{C}_{\text{CRT}}(P_{Y|X_1,X_2})$, with classical CR between the transmitters. 
The green square is the capacity region $\mathcal{C}_{\text{ET}}(P_{Y|X_1,X_2})$, with entangled transmitters. The red area represents rate pairs that can be achieved with entanglement, but cannot be achieved with classical CR.
}
\label{fig:MS_Regions}
\end{figure*}

\subsection{Pseudo-Telepathy Examples}
\label{Subsection:Magic_Example}
As an example, we consider two 
channels that were introduced by Leditzky et al.  \cite{LeditzkyAlhejjiLevinSmith:20p}.

\begin{example}[Magic-square channel]
\label{Example:Magic_Square}
The channel is defined in terms of a pseudo-telepathy game, i.e., a non-local game such that quantum strategies outperform classical strategies and guarantee winning with certainty. 
In the magic square game,  a referee selects one out of nine cells uniformly at random.
Suppose that the referee selected $(r,c)$.
Then, the referee informs Player 1 of the row index $r$, and Player 2 of the column index $c$.
Each player fills three bits in the respective row $r$ and column $c$. 
In order to win the game, they need to satisfy three requirements: they agree on the bit value in $(r,c)$, the row $r$ has even parity, and the column $c$ has odd parity. 
Tables \ref{Table:Magic_Classical}
and \ref{Table:Magic_Quantum}
demonstrate 
 a classical strategy and a quantum strategy, respectively.
If the game is limited to a classical deterministic strategy, then winning is impossible. An attempt towards winning the game is shown in Table~\ref{Table:Magic_Classical}. 
Furthermore, using randomized strategies, the probability of winning is at most $\frac{8}{9}$.
On the other hand, it can be shown that the following quantum strategy wins the game with probability 1:
\begin{itemize}
\item
Before the game begins,
prepare the state
\begin{align}
\ket{\phi_{A_1' A_1'' A_2' A_2''}}=
\frac{1}{2}\left(
\ket{00} \ket{11}+\ket{11} \ket{00}- \ket{01} \ket{10} - \ket{10} \ket{01} \right) \,.
\label{eq:Magic_A1A2}
\end{align}
Send $A_1',A_1''$ to Player 1, and $A_2',A_2''$ to Player 2.  

\item
Play the game using the following strategy.
Having received the row and column indices, each player measures the observables in Table~\ref{Table:Magic_Quantum} simultaneously, and inserts the measurement outcomes into the corresponding cells, where
$\mathsf{X}$, $\mathsf{Y}$, and $\mathsf{Z}$ are the Pauli operators.
 The observables can be measured simultaneously because the three operators in each row and each column commute.
For example, if the referee selected $r=1$ and $c=2$, then Player 1 measures $\mathsf{X}\otimes\identity$, $\mathsf{X}\otimes \mathsf{X}$, and $\identity\otimes \mathsf{X}$, whereas Player 2 measures $\mathsf{X}\otimes \mathsf{X}$, $\mathsf{Y}\otimes \mathsf{Y}$, and $\mathsf{Z}\otimes \mathsf{Z}$.

\end{itemize}
Leditzky et al.  \cite{LeditzkyAlhejjiLevinSmith:20p} defined a MAC such that the channel is ideal for input strategies that win the game, and pure-noise otherwise. 
The precise definition is given below.

In the general description of the non-local game, a referee selects two questions $q_1\in\mathcal{Q}_1$ and $q_2\in\mathcal{Q}_2$ uniformly at random, and
sends the respective question to each player.
The players choose their respective answers $a_1\in\mathcal{A}_1$ and $a_2\in\mathcal{A}_2$ using either deterministic or random strategies $f_k:\mathcal{Q}_k\to \mathcal{A}_k$, for $k=1,2$.
The game is won if $(q_1,q_2,a_1,a_2)\in\mathscr{G}$, where $\mathscr{G}$ is the winning set.
The CHSH game and magic square game are special cases of this description.
Leditzky et al.  \cite{LeditzkyAlhejjiLevinSmith:20p} defined a classical MAC $P_{Y|X_1,X_2}$, with
\begin{align}
\Xset_k &=\mathcal{Q}_k\times \mathcal{A}_k \,,\; k=1,2\\
\Yset&= \mathcal{Q}_1\times \mathcal{Q}_2
\intertext{such that}
P_{Y|X_1,X_2}\big( (\hat{q}_1,\hat{q}_2) \,\big|  q_1,a_1,q_2,a_2  \big)&=
\begin{cases}
\delta_{q_1, \hat{q}_1} \delta_{q_2, \hat{q}_2} &\text{ if } (q_1,q_2,a_1,a_2)\in\mathscr{G}\,,\\
\frac{1}{|\mathcal{Q}_1| |\mathcal{Q}_2|} &\text{otherwise.}
\end{cases}
\label{Equation:MS_channel}
\end{align}
In words, if the inputs $X_1=(q_1,a_1)$ and $X_2=(q_2,a_2)$ win the game, then the decoder receives the question pair precisely, i.e.,
$Y=(q_1,q_2)$ with probability $1$. Otherwise, if the game is lost, then the output $Y$ is uniformly distributed over the question set.

Formally, the magic square game is specified by questions $(q_1,q_2)$ from $\{1,2,3\}\times \{1,2,3\}$,
answers $a_1,a_2$ from $\Aset_1=\Aset_2=\{ 0,1 \}^3$, and the winning set $\mathscr{G}=\big\{$ $\big( q_1,q_2, 
(a_1[j],a_2[j])_{j=1,2,3} \big)
\in\Qset_1\times\Qset_2\times\Aset_1\times\Aset_2:$ $a_1[q_1]=a_2[q_2]$,
 $ a_1[1]+ a_1[2] + a_1[3] \mod 2=0 $, and
 $a_2[1]+ a_2[2] + a_2[3] \mod 2=1$
$\big\}$. Seshadri et al. \cite{SeshadriLeditzkySiddhuSmith:22c} showed that  without assisting resources and with a message-average error criterion, the sum-rate  is bounded by 
$
R_1+R_2\leq 3.02 
$. 
Therefore, this holds for classical CR between the transmitters as well 
by Theorem~\ref{theo:crtMAC_C}, for either  maximal or message-average error criterion (see Remark~\ref{Remark:Time_Sharing}). 
That is, the capacity region with classical CR between the transmitters satisfies
\begin{align}
\mathcal{C}_{\text{CRT}}(P_{Y|X_1,X_2})\subseteq
\{(R_1,R_2):R_1+R_2\leq 3.02\} \,.
\end{align}
The outer bound is depicted in Figure~\ref{fig:MS_Regions} as a gray triangle.

Based on \cite{LeditzkyAlhejjiLevinSmith:20p}, the sum rate $R_1+R_2=2\log(3)\approx 3.17$ is achievable with entangled transmitters.
This can also be obtained as a consequence of our result.
By Theorem~\ref{theo:etMAC_In}, the capacity region with entangled transmitters is given by 
\begin{align}
\mathcal{C}_{\text{ET}}(P_{Y|X_1,X_2})=
\left\{ \begin{array}{rl}
  (R_1,R_2) \,:\;
	&R_1 \leq \log(3)  \\
  &R_2 \leq \log(3) 
	\end{array}
\right\} \,.
\label{eq:Cet_Magic}
\end{align}
The capacity region of the MAC with entangled transmitters is 
the square region in Figure~\ref{fig:MS_Regions}.
The figure demonstrates the violation of the classical upper bound using entanglement, as 
the red area  represents rate pairs that can be achieved with entanglement, but cannot be achieved with classical CR.

 The converse part is immediate since 
the set on the right-hand side of (\ref{eq:Cet_Magic}) is the capacity region of the noiseless MAC
$\widetilde{P}_{Y|X_1,X_2}\big( (\hat{q}_1,\hat{q}_2) \,\big|  q_1,a_1,q_2,a_2  \big)=\delta_{q_1, \hat{q}_1} \delta_{q_2, \hat{q}_2}$ (with or without entanglement resources).
As for the direct part, consider the region in (\ref{eq:inRetx_In}). We choose a bipartite state, classical variables, and POVMs as follows. Let the bipartite state $\ket{\phi_{A_1 A_2}}$ be as in (\ref{eq:Magic_A1A2}), where
$A_k\equiv A_k' A_k''$ for $k=1,2$.
Furthremore, we let $\Vset_0=\emptyset$ and set the  distribution $p_{V_k}$ to be uniform over $\Vset_k=\{1,2,3\}$, for $k=1,2$.
Thus, the random pair $(V_1,V_2)$ is distributed as the referee's questions.
Given $V_k$, Alice $k$ performs a measurement $\Lset_k(V_k)$ as follows. Alice 1 measures the observables in row number $r=V_1$ in Table~\ref{Table:Magic_Quantum},  obtains a random triplet
$W_1\equiv (a_1[j])_{j=1,2,3}$, and transmits $X_1=(V_1,W_1)$.
Similarly, Alice 2 measures the observables in column number $c=V_2$,  obtains 
$W_2\equiv (a_2[j])_{j=1,2,3}$, and transmits $X_2=(V_2,W_2)$. Since $(V_1,V_2,W_1,W_2)$ win the game, we have $Y=(V_1,V_2)$  with probability 1, hence
$I(V_1 V_2;Y)=H(V_1 V_2)=2\log(3)$, $I(V_1;Y|V_2)=H(V_1)=\log(3)$, and $I(V_2;Y|V_1)=H(V_2)=\log(3)$.
This requires an entanglement rate of 
$\theta_E=2$ qubit pairs per transmission.
\end{example}

\begin{example}[Slofstra-Vidick channel]
\label{Example:Slofstra_Vidick}
Another example by Leditzky et al.  \cite{LeditzkyAlhejjiLevinSmith:20p} is  defined in terms of the following non-local game.
Consider a linear system of equations, $\mathbf{H}\mathbf{x}=\mathbf{b}$, over $\mathrm{GF}(2)$, with fixed $\mathbf{H}$ and $\mathbf{b}$, where $\mathbf{H}$ is a  
 binary matrix of size $K\times N$, and $\mathbf{b}$ is a binary vector of length $K$.
That is, the system consists of 
$K$ equations of the form
\begin{align}
\mathbf{h}_k\cdot\,\mathbf{x}=
b_k 
\label{Equation:k_row}
\end{align}
for $k\in \{1,\ldots,K\}$,
where $\mathbf{h}_k$ is the $k$th row in $\mathbf{H}$.

In the linear-system game \cite{SlofstraVidick:18p},  a referee selects two indices, an equation index 
$k\in \{1,\ldots,K\}$ and a variable index $j\in\{ 1,\ldots,N \}$, 
uniformly at random.
Then, the referee informs Player 1 of the index $k$, and Player 2 of the index $j$.
Player 1 responds with a vector
$\mathbf{a}_1 \in \{0,1\}^N$, which is interpreted as an assignment for 
$\mathbf{x}$,
 and  Player 2 responds with a single  bit
$a_2\in\{0,1\}$ that is interpreted as the value of $\mathbf{x}[j]$.
In order to win the game, they need to satisfy two requirements:
\begin{enumerate}[1)]
\item
The response of Player 1 satisfies the $k$th equation. That is, (\ref{Equation:k_row}) holds for
$\mathbf{x}=\mathbf{a}_1$.

\item
Their responses satisfy
$\mathbf{H}[k,j]\cdot(\mathbf{a}_1[j]-a_2)=0$.
That is, either the $j$th variable is not included in the $k$th equation
(i.e., $\mathbf{H}[k,j]=0$), or the players must agree on the value of 
$\mathbf{x}[j]$ (namely, $\mathbf{a}_1[j]=a_2$).
\end{enumerate}
Suppose that the players share entangled systems of dimension $d_E$.
Slofstra and Vidick \cite{SlofstraVidick:18p} showed that quantum strategies outperform classical strategies in this game.
However, the minimal entanglement dimension $d_{E,\min}$ that is required 
in order to win the game with a probability of 
$1-e^{-T}$ is bounded by
\begin{align}
Ce^{T/6}%
\leq d_{E,\min}\leq
C' e^{T/2} %
\label{Equation:SV_dimension}
\end{align}
for all $T>0$, where $C,C'$ are positive constants. See further details in \cite[Th. 1.1]{SlofstraVidick:18p}.
It follows that the game can be won with certainty for
$d_E\to \infty$, and cannot be won with certainty if the entanglement dimension is bounded.

Leditzky et al.  \cite{LeditzkyAlhejjiLevinSmith:20p} considered a classical MAC $P_{Y|X_1,X_2}$ defined as in
(\ref{Equation:MS_channel}), where $\mathscr{G}$ is the winning set for the Slofstra-Vidick game \cite{SlofstraVidick:18p}, as described above,
where $\mathcal{Q}_1=\{1,\ldots,K\}$,
$\mathcal{Q}_2=\{1,\ldots,N\}$,
$\mathcal{A}_1=\{0,1\}^N$, and
$\mathcal{A}_2=\{0,1\}$.
As before, if the inputs $X_1=(q_1,\mathbf{a}_1)$ and $X_2=(q_2,a_2)$ win the game, then the decoder receives the question pair precisely, i.e.,
$Y=(q_1,q_2)$ with probability $1$. Otherwise, if the game is lost, then the output $Y$ is uniformly distributed over the question set.
The capacity region of the classical MAC with unlimited entanglement resources between the transmitters is given by  \cite{LeditzkyAlhejjiLevinSmith:20p}
\begin{align}
\mathcal{C}_{\text{ET}}(P_{Y|X_1,X_2})=
\left\{ \begin{array}{rl}
  (R_1,R_2) \,:\;
	&R_1 \leq \log(K)  \\
  &R_2 \leq \log(N) 
	\end{array}
\right\} \,.
\label{eq:Cet_SV}
\end{align}
This result can also be obtained from Theorem~\ref{theo:etMAC_In} following similar arguments as in Example~\ref{Example:Magic_Square}.

Furthermore, we obtain an achievable rate region with entanglement resources at a limited rate.
By Corollary~\ref{coro:etMAC_Limited}, we have that a rate pair $(R_1,R_2)$ is achievable given entanglement resources at a limited rate of 
$\theta_E=\frac{1}{2}T+\log(C')$, if
\begin{align}
R_1 &\leq \left( 1-2e^{-T} \right)\log(K)-2h_2(
e^{-T}
) \,,
\\
R_2 &\leq \left( 1-2e^{-T} \right)\log(N)-2h_2(
e^{-T}
)
\end{align}
for all $T>0$,
where $h_2(p)=-p\log(p)-(1-p)\log(1-p)$ is the binary entropy function over $(0,1)$.
To show this, we use the
Slofstra-Vidick bound in
(\ref{Equation:SV_dimension}), along with entropy continuity bounds
(see 
\cite{AlhejjiSmith:20c} \cite[Sec. 11.10]{Wilde:17b}).

\end{example}

\section{Main Results --- Entangled and Conferencing Transmitters}
\label{Section:Main_Results_Conferencing}

\subsection{%
Classical Conferencing}
\label{Section:Conferencing_Results}
In this section, we consider the combination of entanglement resources and classical conferencing between the transmitters (see Figure~\ref{fig:MentangledTx_Conferencing}).
In addition to pre-shared entanglement resources,  %
the transmitters can now communicate with each other over noise-free links at fixed rates, with a classical capacity of $C_{12}$ from Transmitter 1 to Transmitter 2, and  
$C_{21}$ from Transmitter 2 to Transmitter 1. 

 Define the following rate region,
\begin{align}%
\mathcal{R}_{\text{ET-C}}(P_{Y|X_1,X_2},C_{12},C_{21})=
\bigcup_{ p_{V_0} p_{V_1|V_0} p_{V_2|V_0} \,,\; \varphi_{A_1 A_2} \,,\; \Lset_1 \otimes \Lset_2 }
\left\{ \begin{array}{rl}
  (R_1,R_2) \,:\;
	R_1 &\leq I(V_1;Y|V_0 V_2)+C_{12}  \\
  R_2 &\leq I(V_2;Y|V_0 V_1)+C_{21} \\
	R_1+R_2 &\leq I(V_1 V_2;Y|V_0)+C_{12}+C_{21}\\
        R_1+R_2 &\leq I(V_1 V_2;Y)\\
	\end{array}
\right\} \,.
\label{eq:inRetxD_Conferencing}
\end{align}%
The union on the right-hand side of (\ref{eq:inRetxD_Conferencing}) is the same as in
(\ref{eq:inRetx_In}).
Similarly, we define the region $\mathcal{R}_{\text{ET-C}}(P_{Y|X_1,X_2},C_{12},C_{21})$ with a union over all joint distributions $p_{V_0, V_1, V_2}$.

\begin{theorem}
\label{theo:etMAC_Conferencing}
The capacity region of a classical MAC $P_{Y|X_1,X_2}$ with entangled  transmitters and classical conferencing satisfies
\begin{align}
\mathcal{C}_{\text{ET-C}}(P_{Y|X_1,X_2},C_{12},C_{21})\supseteq \mathcal{R}_{\text{ET-C}}(P_{Y|X_1,X_2},C_{12},C_{21}) \,,
\intertext{and}
\mathcal{C}_{\text{ET-C}}(P_{Y|X_1,X_2},C_{12},C_{21})\subseteq \mathcal{O}_{\text{ET-C}}(P_{Y|X_1,X_2},C_{12},C_{21}) \,.
\end{align}
Furthermore, the capacity region is given by
\begin{align}
\mathcal{C}_{\text{ET-C}}(P_{Y|X_1,X_2},C_{12},C_{21})=
\bigcup_{\ell=1}^{\infty}\frac{1}{\ell}
\mathcal{R}_{\text{ET-C}} (P_{Y|X_1,X_2}^\ell,C_{12},C_{21}) \,,
\end{align}
\end{theorem}
The proof of Theorem~\ref{theo:etMAC_Conferencing} is given in Section%
~\ref{app:etMAC_Conferencing}.

\begin{remark}
If the conferencing rates are sufficiently high, then we obtain the full-cooperation region,
\begin{align}%
\mathcal{C}_{\text{ET-C}}(P_{Y|X_1,X_2},C_{12},C_{21})=
\left\{ 
  (R_1,R_2) \,:\;
	R_1+R_2 \leq C_{\text{joint}}
\right\} 
\label{eq:inRetx_Conferencing}
\end{align}
where
$C_{\text{joint}}\equiv \max_{p_{X_1 X_2}} I(X_1 X_2;Y) $.
In particular, this holds if both conferencing rates are higher than $C_{\text{joint}}$. In this case, we effectively have a single-user channel from $ (X_1,X_2)$ to $Y$. On the other hand, if the conferencing rates are zero, i.e., $C_{12}=C_{21}=0$,
then we recover the achievable region without conferencing, as in Theorem~\ref{theo:etMAC_In}. That is, 
$
\mathcal{C}_{\text{ET-C}}(P_{Y|X_1,X_2},0,0)=
\mathcal{C}_{\text{ET}}(P_{Y|X_1,X_2})
$.
\end{remark}

\begin{remark}
We have compared the effect of entanglement and CR on communication performance in 
Remark~\ref{Remark:Common_Randomness}.
Both entanglement and CR are static resources of \emph{non-signaling correlation}, which cannot be used in order to send information. Conferencing, however, is a \emph{dynamic} resource of cooperation,  which can be used by itself to send information from one transmitter to the other.
In this sense, conferencing is stronger than entanglement. Thereby, one may ask, for example, is conferencing at a low rate better than entanglement at a high rate?
While this question may seem unfair, it can be justified by the delay that is caused by using conferencing.
Based on Theorem~\ref{theo:etMAC_Conferencing}, the answer is no. It can be easily seen that when the values of $C_{12}$ and $C_{21}$ are very low, the rate increase is negligible as well. Whereas, we have seen that sufficient entanglement without conferencing can lead to a significant increase.
\end{remark}

\begin{remark}
Conferencing is achieved through interaction between the two transmitters. As already mentioned, we only consider trustworthy senders here. For future communication systems, especially for 6G, it is also very important to examine communication protocols for non-trustworthy communication parties. The complexity theory shows that performance behavior, which is also very interesting for such communication parties, can be increased. In the last 40 years, enormous progress has been made in complexity theory, which should also be of interest for communication systems. In particular, the work for prover and verifier interaction is very relevant. Techniques from the coding theory were also often used here in finding of corresponding protocols. For example, it was shown in \cite{Z7} that through interaction between a prover and a verifier with polynomial time, and the verifier using randomness exactly then all languages from PSPACE (polynomial space Turing machines) are decidable. In an interaction between two provers and a verifier with polynomial time, all languages from NEXP, i.e. Turing machines with non-deterministic exponential computing time, can be decided using randomness \cite{Z8}. If the performance of the verifier is described as a poly-logarithmic use of random bits with a constant number of information bits greater than or equal to 3, then in an optimal interaction with a prover the verifier can already decide all languages from NP, i.e. non- deterministic polynomial time \cite{Z9, Z10}. Of course, much more complex communication scenarios occur in real communication systems than those considered in 
\cite{Z7,Z8,Z9,Z10}, but this research direction is nevertheless particularly interesting with regard to trustworthiness for 6G. It is fascinating to see that one of the central problems in mathematics, (Millennium prize \cite{Z12}), can be represented in an equivalent form by a simple interactive communication task (see \cite{Z11,Z12,Z13} for an overview and status of the problems).
For the case with entanglement assistance, this is open. Another very interesting property of interactive communication in complexity theory is: In the communication scenario with one prover and two verifiers, there is a zero-knowledge interactive proof if and only if a one-way function exists \cite{Z19,Z20}. It should be noted here that the existence of a one-way function as a central cryptographic primitive cannot be proved so far.  
In complexity theory, also for certain interactive communication scenarios, the type of interactivity, that is, how many rounds of interaction are necessary to solve certain tasks, is interesting. In \cite{Z7,Z8} there is no bound on the number of rounds of interaction necessary. In \cite{CAI} it has now been shown for two prover and two verifier interactive systems that any language from PSPACE can be decided with a one round protocol. For the MAC with conferencing encoders and also the AV-MAC with conferencing encoders, one round of conferencing is sufficient for each. On the other hand, the existence of interactive zero-knowledge proofs for the two prover and one verifier scenario can be directly proved \cite{Z21}. That is, no cryptographic assumptions are needed. This is another interesting example of the special potential of interactive and cooperative communication scenarios.

\end{remark}

\subsection{%
Quantum Conferencing}
\label{Section:Conferencing_Results_Q}
In this section, we consider the MAC with  entanglement and quantum conferencing between the transmitters (see Figure~\ref{fig:MentangledTx_Conferencing_Q}).
The transmitters can now communicate with each other over noise-free quantum links at fixed rates, with a quantum capacity of $Q_{12}$ from Transmitter 1 to Transmitter 2, and  
$Q_{21}$ from Transmitter 2 to Transmitter 1. Furthermore, the transmitters share entanglement resources that can be used at both the conferencing and the encoding stages.
 We establish an inner bound on the capacity region. %
 Define 
\begin{align}%
\mathcal{R}_{\text{ET-Q}}(P_{Y|X_1,X_2},Q_{12},Q_{21})=
\bigcup_{ p_{V_0} p_{V_1|V_0} p_{V_2|V_0} \,,\; \varphi_{A_1 A_2} \,,\; \Lset_1 \otimes \Lset_2 }
\left\{ \begin{array}{rl}
  (R_1,R_2) \,:\;
	R_1 &\leq I(V_1;Y|V_0 V_2)+2Q_{12}  \\
  R_2 &\leq I(V_2;Y|V_0 V_1)+2Q_{21} \\
	R_1+R_2 &\leq I(V_1 V_2;Y|V_0)+2(Q_{12}+Q_{21})\\
        R_1+R_2 &\leq I(V_1 V_2;Y)\\
	\end{array}
\right\} \,.
\label{eq:inRetxD_Conferencing_Q}
\end{align}%
The union on the right-hand side of (\ref{eq:inRetxD_Conferencing}) is the same as in
(\ref{eq:inRetx_In}). 

\begin{theorem}
\label{theo:etMAC_Conferencing_Q}
The capacity region of a classical MAC $P_{Y|X_1,X_2}$ with entangled  transmitters and quantum conferencing satisfies
\begin{align}
\mathcal{C}_{\text{ET-Q}}(P_{Y|X_1,X_2},C_{12},C_{21})\supseteq \mathcal{R}_{\text{ET-Q}}(P_{Y|X_1,X_2},C_{12},C_{21}) \,.
\end{align}
\end{theorem}
The  proof follows similar lines as for Theorem~\ref{theo:etMAC_Conferencing} in Appendix~\ref{app:etMAC_Conferencing}. Here, however, each encoder uses superdense coding in order to increase the rate of her conferencing message. Alice 1 sends $nR_{1}'$ bits to Alice 2 through a quantum transmission of $\frac{1}{2} nR_1'$ qubits via her quantum conferencing link, using   $\frac{1}{2}n R_1'$ entangled pairs. Similarly, Alice 2 sends $nR_{2}'$ bits to Alice 2 by sending $\frac{1}{2} nR_2'$ qubits  along with $\frac{1}{2}n R_2'$ entangled pairs. 
 This is permissible if the common information rates are each bounded by \emph{twice} the respective conferencing capacity, i.e.,
 $R_1' \leq 2Q_{12}$ and $R_2' \leq 2Q_{21}$.
The rest of the proof is identical to that of Theorem~\ref{theo:etMAC_Conferencing}. Hence, The details are omitted.

\begin{figure*}[tb]
\includegraphics[scale=1.85,trim={2cm 21cm 2cm 
4.5cm},clip]{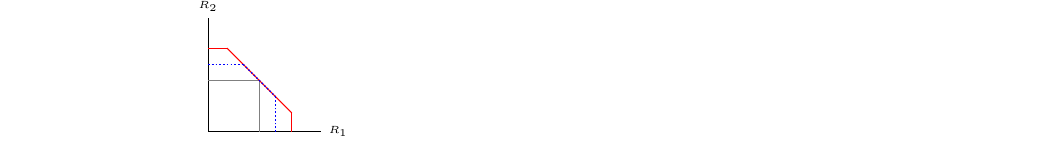} %
\caption{Achievable rate regions for the magic-square  multiple-access channel $P_{Y|X_1,X_2}$  with entangled transmitters and, either classical or quantum, conferencing, as in Example~\ref{Example:Magic_Square}.
The inner square indicates the capacity region $\mathcal{C}_{\text{ET}}(P_{Y|X_1,X_2})$ with entangled transmitters. 
The pentagon below the blue dashed line is an achievable region for the MAC with entangled transmitters and classical conferencing, for 
$C_{12}=C_{21}=0.5$ (see \eqref{eq:Cet_Magic_C}).
The pentagon below the red solid line is achievable when the conferencing links are quantum as well, for 
$Q_{12}=Q_{21}=0.5$ (see \eqref{eq:Cet_Magic_Q}).%
}
\label{fig:MS_Regions_C}
\end{figure*}

\subsection{Example}
\label{Subsection:Magic_Example_Conferencing}
To demonstrate our results on the classical MAC with entangled transmitters and conferencing,
we revisit the magic-square channel from Subsection~\ref{Subsection:Magic_Example}.

\addtocounter{example}{-2}
\begin{example}[Continued]
\label{Example:Magic_Square_Conferencing}
Consider the magic-square MAC from Example~\ref{Example:Magic_Square}, where the channel output is a noiseless copy of the referee's chosen  cell if the inputs win the game, and it is pure noise otherwise.

Figure~\ref{fig:MS_Regions_C} shows achievable rate regions in the three settings that we have considered.
We have seen that 
since the game can be won with \mbox{probability $1$} given entanglement resources, the capacity region with entangled transmitters, without conferencing, is given by
\cite{LeditzkyAlhejjiLevinSmith:20p}
\begin{align}
\mathcal{C}_{\text{ET}}(P_{Y|X_1,X_2})=
\left\{ \begin{array}{rl}
  (R_1,R_2) \,:\;
	&R_1 \leq \log(3)  \\
  &R_2 \leq \log(3) 
	\end{array}
\right\} \,.
\label{eq:Cet_Magic_No_Conferencing}
\end{align}
The capacity region of the MAC with entangled transmitters and without conferencing is 
the square region in Figure~\ref{fig:MS_Regions_C}.
%

Now, consider conferencing of classical messages between the transmitters, at rates $C_{12}=C_{21}=0.5$.
In this case, we obtain the following achievable region 
\begin{align}
\mathcal{R}_{\text{ET-C}}(P_{Y|X_1,X_2})\supseteq
\left\{ \begin{array}{rrl}
  (R_1,R_2) \,:\;&
	R_1 &\leq \log(3)+0.5  \\
  &R_2 &\leq \log(3)+0.5\\ 
  &R_1+R_2 &\leq 2\log(3)
	\end{array}
\right\} \,.
\label{eq:Cet_Magic_C}
\end{align}
If the conferencing links are quantum, however, at the same rates $Q_{12}=Q_{21}=0.5$, then we have an even larger achievable region,
\begin{align}
\mathcal{R}_{\text{ET-Q}}(P_{Y|X_1,X_2})\supseteq
\left\{ \begin{array}{rrl}
  (R_1,R_2) \,:\;&
	R_1 &\leq \log(3)+1  \\
&  R_2 &\leq \log(3)+1\\ 
&  R_1+R_2 &\leq 2\log(3)
	\end{array}
\right\} \,.
\label{eq:Cet_Magic_Q}
\end{align}
The bounds are depicted in Figure~\ref{fig:MS_Regions_C}. The pentagons, below the blue dashed line and below the red solid line, indicate the achievable regions on the right hand side of  \eqref{eq:Cet_Magic_C} and \eqref{eq:Cet_Magic_Q}, respectively.
We show 
achievability based on Theorem~\ref{theo:etMAC_Conferencing} and Theorem~\ref{theo:etMAC_Conferencing_Q}, respectively,
using the same choice of a bipartite state, classical variables, and POVMs, as in the first part of Example~\ref{Example:Magic_Square}. Specifically, let the bipartite state $\ket{\phi_{A_1 A_2}}$ be as in (\ref{eq:Magic_A1A2}), where
$A_k\equiv A_k' A_k''$ for $k=1,2$.
Furthermore, we let $\Vset_0=\emptyset$ and set the  distribution $p_{V_k}$ to be uniform over $\Vset_k=\{1,2,3\}$, for $k=1,2$, as the referee's questions.
Given $V_1$, 
Alice 1 measures the observables in row number $r=V_1$ in Table~\ref{Table:Magic_Quantum},  obtains a random triplet
$W_1\equiv (a_1[j])_{j=1,2,3}$, and transmits $X_1=(V_1,W_1)$.
Similarly, Alice 2 measures the observables in column number $c=V_2$,  obtains 
$W_2\equiv (a_2[j])_{j=1,2,3}$, and transmits $X_2=(V_2,W_2)$. Since $(V_1,V_2,W_1,W_2)$ win the game, we have $Y=(V_1,V_2)$  with probability 1, hence
$I(V_1 V_2;Y)=H(V_1 V_2)=2\log(3)$, $I(V_1;Y|V_2)=H(V_1)=\log(3)$, and $I(V_2;Y|V_1)=H(V_2)=\log(3)$.
\end{example}

\section{Purification and Cardinality}
In this section, we 
show that the inner and outer regions 
can be exhausted with pure states and bounded cardinality.

\subsection{Inner Bound}
\label{app:CardCl}
We prove Lemma~\ref{lemm:CardCl} and show that
the region $\mathcal{R}_{\text{ET}}(P_{Y|X_1,X_2})$ can be exhausted with pure states and bounded cardinality.

\subsubsection{Purification}
First, we consider a given Hilbert space
$\Hset_{A_1}\otimes \Hset_{A_2}$, and
show that the union over bipartite states $\varphi_{A_1 A_2}$ is exhausted by pure states.
Fix a Hilbert space $\Hset_{A_1}\otimes \Hset_{A_2}$.
Consider a given distribution $\{ p_{V_1|V_0}(\cdot|v_0)p_{V_2|V_0}(\cdot|v_0)\}$, a bipartite state 
$\varphi_{A_1 A_2}$, and measurements $\Lset_1(v_0,v_1)$ and $\Lset_2(v_0,v_2)$ on $A_1$ and $A_2$, respectively.
Let $\mathfrak{R}(\varphi,\Lset_1,\Lset_2)$ denote the associated rate region,
\begin{align}%
\mathfrak{R}(\varphi,\Lset_1,\Lset_2)=
\left\{ \begin{array}{rl}
  (R_1,R_2) \,:\;
	R_1 &\leq I(V_1;Y|V_0 V_2)  \\
  R_2 &\leq I(V_2;Y|V_0 V_1) \\
	R_1+R_2 &\leq I(V_1 V_2;Y|V_0)
	\end{array}
\right\} \,.
\label{eq:inRphi}
\end{align}%
Given a joint state $\varphi_{A_1 A_2}$,
consider a spectral decomposition,
\begin{align}
\varphi_{A_1 A_2}=\sum_{z\in\Zset} p_Z(z) \ketbra{ \phi_z }_{A_1 A_2} \,,
\label{eq:Cardinality_Spectral}
\end{align}
where $p_Z$ is a probability distribution, such that the pure states, $\ket{\phi_z}_{A_1 A_2}$, $z\in\Zset$, form an orthonormal basis for 
$\Hset_{A_1} \otimes \Hset_{A_2}$.
Then, $\varphi_{A_1 A_2}$ has the following purification,
\begin{align}
\ket{\psi_{A_1 A_2 E_1 E_2}}= \sum_{z\in\Zset} \sqrt{p_Z(z)} \ket{ \phi_z }_{A_1 A_2}\otimes \ket{z}_{E_1} \otimes \ket{z}_{E_2}
\,.
\end{align}
Now, we consider rate region $\mathfrak{R}(\psi,\Lset_1',\Lset_2')$ that is associated with the following choice of state, distribution, and measurements.
Set the distribution $\{ p_{V_1|V_0}(\cdot|v_0)p_{V_2|V_0}(\cdot|v_0)\}$ as before. Let
Alice 1 and Alice 2 share the state $\ket{\psi_{A_1 E_1 A_2  E_2}}$, and 
suppose that Alice 1 performs a measurement on $(A_1,E_1)$, and Alice 2 on $(A_2,E_2)$, with the following POVMs,
\begin{align}
L_1'(x_1,z|v_0,v_1)&= L(x_1|v_0,v_1)\otimes \ketbra{z}_{E_1} \,,\\
L_2'(x_2,z|v_0,v_2)&= L(x_2|v_0,v_2)\otimes \ketbra{z}_{E_2} \,,
\end{align}
for $(v_0,v_1,v_2,x_1,x_2,z)\in \Vset_0\times\Vset_1\times\Vset_2\times\Xset_1\times\Xset_2\times \Zset$.
The corresponding input distribution $p'_{X_1,X_2|V_0,V_1,V_2}$ is given by
\begin{align}
p'_{X_1,X_2|V_0,V_1,V_2}(x_1,x_2|v_0,v_1,v_2)&=
\sum_{z\in\Zset} p_Z(z) \trace\left[ \left( L_1(x_1|v_0,v_1)\otimes L_2(x_2|v_0,v_2) \right) \ketbra{ \phi_z }_{A_1 A_2}  \right]
\\
&=
 \trace\left[ \left( L_1(x_1|v_0,v_1)\otimes L_2(x_2|v_0,v_2) \right)\left(\sum_{z\in\Zset}  p_Z(z)\ketbra{ \phi_z }_{A_1 A_2} \right) \right]
\\
&=
 \trace\left[ \left( L_1(x_1|v_0,v_1)\otimes L_2(x_2|v_0,v_2) \right)\varphi_{A_1 A_2} \right]
\\
&=
p_{X_1,X_2|V_0,V_1,V_2}(x_1,x_2|v_0,v_1,v_2)
\end{align}
where the third equality follows from (\ref{eq:Cardinality_Spectral}), and the last from (\ref{eq:inRsc_Distribution}).
Therefore, $\mathfrak{R}(\psi,\Lset_1',\Lset_2')=\mathfrak{R}(\varphi,\Lset_1,\Lset_2)$.
We deduce that the entire region $\mathcal{R}_{\text{ET}}(P_{Y|X_1,X_2})$, as in (\ref{eq:inRetx_In}), can be obtained from pure states.

\subsubsection{Cardinality Bounds}
\label{subsection:Cardinality_Proof}
Now, we bound the cardinality of the alphabet $\Vset_0$. Consider a given distribution $\{ p_{V_1|V_0}(\cdot|v_0)p_{V_2|V_0}(\cdot|v_0)\}$, a joint state 
$\varphi_{A_1 A_2}$, and POVM collections, $\Lset_1(v_0,v_1)$ and $\Lset_2(v_0,v_2)$.
Define a map $\tau_0:\Vset_0\rightarrow \mathbb{R}^{3}$ by
\begin{align}%
\tau_{0}(v_0)= \Big(  %
   I(V_1;Y|V_2, V_0=v_0)   \,,\; I(V_2;Y|V_1, V_0=v_0)  \,,\; I(V_1, V_2;Y|V_0=v_0)  \Big) \,.
\end{align}%
The map $\tau_0$ can be extended to a map that  acts on probability distributions as follows,
\begin{align}%
T_{0} \,:\; p_{V_0}(\cdot)  \mapsto
\sum_{v_0\in\Vset_0} p_{V_0}(v_0) \tau_{0}(v_0)= \Big(  I(V_1;Y|V_0 V_2) \,,\; I(V_2;Y|V_0 V_1)   \,,\; I(V_1 V_2;Y|V_0 )   \Big)  \,.
\end{align}%
According to the Fenchel-Eggleston-Carath\'eodory theorem \cite{Eggleston:66p}, any point in the convex closure of a connected compact set within $\mathbb{R}^d$ belongs to the convex hull of $d$ points in the set. 
Since the map $T_{0}$ is linear, it maps  the set of distributions on $\Vset_0$ to a connected compact set in $\mathbb{R}^{3}$. %
Thus, for every  $p_{V_0}$, 
there exists a probability distribution $p_{\bar{V}_0}$ on a subset $\bar{\Vset}_0\subseteq \Vset_0$ of size $%
3$, such that 
$%
T_{0}(p_{{\bar{V}_0}})=T_{0}(p_{V_0}) %
$. %
We deduce that alphabet size can be restricted to $|\Vset_0|\leq 3$, while preserving $I(V_1;Y|V_0 V_2)$, $ I(V_2;Y|V_0 V_1)$, and $I(V_1 V_2;Y|V_0 )$.

Next, we bound the alphabet size for the auxiliary variables $V_1$.
A standard application of the Fenchel-Eggleston-Carath\'eodory theorem does not work here, as it
results in a cardinality bound of 
$|\mathcal{V}_2|\cdot 3(|\mathcal{X}_1 ||\mathcal{X}_2 |+1$  on $\mathcal{V}_1$.
Thereby, we use the perturbation method
\cite{GohariAnantharam:12p,GraceGuha:22c}.

Fix a joint state $\varphi_{A_1 A_2}$, POVM collections, $V_0=v_0$, and $p_{V_2|V_0}(\cdot|v_0)$.
Since our rate region is a convex set, it can be characterized by determining the supporting Hyperplanes. 
Furthermore,
due to the pentagonal form of each region in \eqref{eq:inRphi}, we may focus on the corner points and consider maximizing the following function,
\begin{align}
G_\lambda(p_{V_1|V_0})=
(1-\lambda) I(V_1;Y|V_2)+\lambda I(V_2;Y)
\label{Equation:G_Lambda}
\end{align}
for $\lambda\in \left[\frac{1}{2},1 \right]$.

To simplify notation, we omit the conditioning on $V_0=v_0$.
Let $(V_1,V_2)\sim 
p_{V_1} p_{V_2}$, and consider the joint distribution
\begin{align}
&Q(v_1,v_2,x_1,x_2,y)\equiv   
p_{V_1}(v_1)p_{V_2}(v_2)
\nonumber\\&\cdot
\trace\left[ (L_1(x_1|v_1)\otimes L_2(x_2|v_2)) \varphi_{A_1 A_2} \right] P_{Y|X_1,X_2}(y|x_1,x_2) \,.
\label{eq:thetaU_Unperturb}
\end{align} 
Then, consider the perturbed distribution,
\begin{align}
Q_\epsilon(v_1)\equiv p_{V_1}(v_1)[1+\epsilon f(v_1)]
\end{align}
and 
\begin{align}
Q_\epsilon(v_1,v_2,x_1,x_2,y)\equiv   
Q_\epsilon(v_1) p_{V_2}(v_2) \trace\left[ (L_1(x_1|v_0,v_1)\otimes L_2(x_2|v_0,v_2)) \varphi_{A_1 A_2} \right] P_{Y|X_1,X_2}(y|x_1,x_2) \,,
\label{eq:thetaU_perturb}
\end{align}
where $f(\cdot)$ are real-valued coefficients that satisfies
$1+\epsilon f(v_1)\geq 0$ for all $v_1\in\mathcal{V}_1$.
We note that $f$ may depend on $v_0$ as well, since we have conditioned on $V_0=v_0$.

We also require that the coefficients $f(\cdot)$ satisfy
\begin{subequations}
\label{Equation:f_Requirements}
\begin{align}
\mathbb{E} f(V_1)&=\sum_{v_1\in\mathcal{V}_1} Q(v_1)f(v_1)=0
\label{Equation:f_E}
\intertext{and}
\mathbb{E} [f(V_1)|X_1=x_1,X_2=x_2]&=\sum_{(v_1,v_2)\in\mathcal{V}_1\times \mathcal{V}_2} Q(v_1,v_2|x_1,x_2)f(v_1)=0 \,,\;\text{for $(x_1,x_2)\in\mathcal{X}_1\times \mathcal{X}_2$} \,.
\label{Equation:f_E_x12}
\end{align}
\end{subequations}
There exists a nonzero solution for this linear equation system, provided that 
\begin{align}
|\mathcal{V}_1|\geq
|\mathcal{X}_1||\mathcal{X}_2|+1
\end{align}
since  \eqref{Equation:f_Requirements}
consists of $
|\mathcal{X}_1||\mathcal{X}_2|+1$ constraints.

Observe that \eqref{Equation:f_E} guarantees that $Q_\epsilon$ forms a probability distribution as well. 
Furthermore, by \eqref{Equation:f_E_x12},
$Q_\epsilon(x_1,x_2)=Q(x_1,x_2)$, hence the marginal distribution of 
$(X_1,X_2,Y)$ is preserved.
Now, let 
$(\bar{V}_1,\bar{V}_2,\bar{X}_1,\bar{X}_2,\bar{Y})\sim Q_\epsilon$ be a random tuple that is distributed according to the perturbed distribution.
Then,
\begin{align}
I(\bar{V}_1;\bar{Y}|\bar{V}_2)&=
H(\bar{V}_1,\bar{V}_2)-H(\bar{Y},\bar{V}_1,\bar{V}_2)
-H(\bar{V}_2)+H(\bar{Y},\bar{V}_2)
\nonumber\\
&=
H(V_1,V_2)-H(Y,V_1,V_2)
-H(V_2)+H(\bar{Y},\bar{V}_2)
+\epsilon[  H_f(V_1,V_2)
- H_f(Y,V_1,V_2)] \,,
\end{align}
and 
\begin{align}
I(\bar{V}_2;\bar{Y})&=
H(Y)+H(V_2)-H(\bar{Y},\bar{V}_2) \,,
\end{align}
where
\begin{align}
&H_f(V_1,V_2)=
-\sum_{v_1,v_2} Q(v_1,v_2)f(v_1)\log Q(v_1,v_2) \,,
\label{Equation:f_entropy_v12}
\\
&H_f(Y,V_1,V_2)=
-\sum_{v_1,v_2,y} Q(v_1,v_2,y)f(v_1)\log Q(v_1,v_2,y) \,.
\label{Equation:f_entropy_v12_y}
\end{align}

If $p_{V_1}\equiv Q_0$ attains the maximum of the function
$G_\lambda$ (see \eqref{Equation:G_Lambda}), then the derivatives satisfy
\begin{align}
\frac{\partial}{\partial \epsilon} [ (1-\lambda) I(\bar{V}_1;\bar{Y}|\bar{V}_2)+\lambda I(\bar{V}_2;\bar{Y})]
\Big|_{\epsilon=0}
&= 0 \,,
\label{Equation:epsilon_1d}
\intertext{and}
\frac{\partial^2}{\partial \epsilon^2} [ (1-\lambda) I(\bar{V}_1;\bar{Y}|\bar{V}_2)+\lambda I(\bar{V}_2;\bar{Y})]
\Big|_{\epsilon=0}
&\leq 0 \,.
\label{Equation:epsilon_2d}
\end{align}
The latter reduces to 
\begin{align}
-(2\lambda-1)
\frac{\partial^2}{\partial \epsilon^2} H(\bar{Y},\bar{V}_2) \Big|_{\epsilon=0}
\leq 0 \,,
\end{align}
which is equivalent to
\begin{align}
\mathbb{E}\left(
[\mathbb{E}(f(V_1)|V_2,Y)]^2
\right)\leq 0 \,,
\end{align}
by \cite[Lemma 2, part 2]{GohariAnantharam:12p}.
In particular, this holds with equality if
\begin{align}
\mathbb{E}(f(V_1)|V_2=v_2,Y=y)=0 \,,
\end{align}
for all $(v_2,y)$ in the support of 
$Q(v_2,y)$, which implies 
\begin{align}
Q_\epsilon(v_2,y)&=Q(v_2,y) \,.
\end{align}
Thus,
\begin{align}
I(\bar{V}_1;\bar{Y}|\bar{V}_2)&=
I(V_1;Y|V_2)
+\epsilon[  H_f(V_1,V_2)
-  H_f(Y,V_1,V_2)]
\,,
\end{align}
and 
\begin{align}
I(\bar{V}_2;\bar{Y})&=
I(V_2;Y) \,.
\end{align}
By \eqref{Equation:epsilon_1d}, we now have
\begin{align}
&
\frac{\partial}{\partial \epsilon} [ (1-\lambda) I(\bar{V}_1;\bar{Y}|\bar{V}_2)+\lambda I(\bar{V}_2;\bar{Y})]
\Big|_{\epsilon=0}
\nonumber\\&
=(1-\lambda)(  H_f(V_1,V_2)
- H_f(Y,V_1,V_2))
\nonumber\\&
= 0 \,.
\end{align}
We deduce that the perturbed distribution attains the same (maximal) value of the function 
$G_\lambda$. Therefore, we can restrict the union to $|\Vset_1|= |\Xset_1| |\Xset_2|+1$.
By symmetry, this holds for 
the cardinality of $V_2$ as well. 
%
%

\subsection{Outer Bound}
\label{app:CardCl_Out}
Next, we show that union for   
the outer bound $\mathcal{O}_{\text{ET}}(P_{Y|X_1,X_2})$ can be restricted as well.
Based on exactly the same arguments, the union can be exhausted with pure states and with 
$|\mathcal{V}_0|\leq 3$.

We bound the alphabet size for the auxiliary variable $V_1$, using similar perturbation arguments.
%
Here, however, we fix
$p_{V_2|V_0,V_1}$ as well.
As before,  consider maximizing the following function,
\begin{align}
\tilde{G}_\lambda(p_{V_1|V_0})=
(1-\lambda) I(V_1;Y|V_2)+\lambda I(V_2;Y)] \,,
\end{align}
for $\lambda\in [0,1]$.

To simplify notation, we omit the conditioning on $V_0=v_0$.
Let $(V_1,V_2)\sim 
p_{V_1} p_{V_2|V_1}$, and consider the joint distribution
\begin{align}
\tilde{Q}(v_1,v_2,x_1,x_2,y)\equiv   
p_{V_1}(v_1)p_{V_2|V_1}(v_2|v_1) \trace\left[ (L_1(x_1|v_1)\otimes L_2(x_2|v_2)) \varphi_{A_1 A_2} \right] P_{Y|X_1,X_2}(y|x_1,x_2) \,.
\label{eq:thetaU_Unperturb_Outer}
\end{align} 
Then, consider the perturbed distribution,
\begin{align}
\tilde{Q}_\epsilon(v_1,v_2)=Q(v_1,v_2)[1+\epsilon \tilde{f}(v_1)] \,,
\end{align}
and
\begin{align}
\tilde{Q}_\epsilon(v_1,v_2,x_1,x_2,y)\equiv   
\tilde{Q}_\epsilon(v_1,v_2) \trace\left[ (L_1(x_1|v_1)\otimes L_2(x_2|v_2)) \varphi_{A_1 A_2} \right] P_{Y|X_1,X_2}(y|x_1,x_2) \,,
\label{eq:thetaU_perturb_Outer}
\end{align}
where $\tilde{f}(\cdot)$ is a real-valued function that satisfies
$1+\epsilon \tilde{f}(v_1)\geq 0$.
We also require that $\tilde{f}$ satisfies the equation system
\eqref{Equation:f_Requirements},
which guarantees that
$\tilde{Q}_\epsilon(x_1,x_2)=Q(x_1,x_2)$.
There exists a nonzero perturbation function that satisfies this system of $|\mathcal{X}_1||\mathcal{X}_2|+1$ linear equations, provided that 
$
|\mathcal{V}_1|\geq
|\mathcal{X}_1||\mathcal{X}_2|+1 
$. 

Hence, for
$(\breve{V}_1,\breve{V}_2,\breve{X}_1,\breve{X}_2,\breve{Y})\sim \tilde{Q}_\epsilon$, 
%
\begin{align}
I(\breve{V}_1;\breve{Y}|\breve{V}_2)&=
H(\breve{V}_1,\breve{V}_2)-H(\breve{Y},\breve{V}_1,\breve{V}_2)
-H(\breve{V}_2)+H(\breve{Y},\breve{V}_2)
\\
&=
H(V_1,V_2)-H(Y,V_1,V_2)
-H(V_2)+H(\breve{Y},\breve{V}_2)
+\epsilon[  H_{\tilde{f}}(V_1,V_2)
-  H_{\tilde{f}}(Y,V_1,V_2)]
\end{align}
and 
\begin{align}
I(\breve{V}_2;\breve{Y})&=
H(\breve{Y})+H(\breve{V}_2)-H(\breve{Y},\breve{V}_2)
\\
&=
H(Y)+H(V_2)-H(\breve{Y},\breve{V}_2)
\end{align}
where the perturbed entropies 
$H_{\tilde{f}}$ are defined as in 
\eqref{Equation:f_entropy_v12}-\eqref{Equation:f_entropy_v12_y}.
If $p_{V_1}\equiv \tilde{Q}_0$ attains the maximum of 
$\tilde{G}_\lambda$, then
\begin{align}
&\frac{\partial^2}{\partial \epsilon^2} [ (1-\lambda) I(\breve{V}_1;\breve{Y}|\breve{V}_2)+\lambda I(\breve{V}_2;\breve{Y})]
\Big|_{\epsilon=0}
\nonumber\\&
=
-(2\lambda-1)\frac{\partial^2}{\partial \epsilon^2} H(\breve{Y},\breve{V}_2) \Big|_{\epsilon=0}
\nonumber\\&
\leq 0 \,,
\end{align}
which is equivalent to
\begin{align}
\mathbb{E}\left(
[\mathbb{E}(\tilde{f}(V_1)|V_2,Y)]^2
\right)\leq 0
\end{align}
by \cite[Lemma 2, part 2]{GohariAnantharam:12p}.
In particular, this holds with equality if
$
\mathbb{E}(\tilde{f}(V_1)|V_2=v_2,Y=y)=0
$, in which case 
\begin{align}
\tilde{Q}_\epsilon(v_2,y)&=\tilde{Q}(v_2,y) \,.
\end{align}
It follows that 
\begin{align}
I(\breve{V}_1;\breve{Y}|\breve{V}_2)&=
I(V_1;Y|V_2)
+\epsilon[  H_{\tilde{f}}(V_1,V_2)
- H_{\tilde{f}}(Y,V_1,V_2)]
\,,
\end{align}
and 
\begin{align}
I(\breve{V}_2;\breve{Y})&=
I(V_2;Y)
\,.
\end{align}
Since $\tilde{Q}_0$ attains the maximum of 
$\tilde{G}_\lambda$, we also have
\begin{align}
&\frac{\partial}{\partial \epsilon} [ (1-\lambda) I(\breve{V}_1;\breve{Y}|\breve{V}_2)+\lambda I(\breve{V}_2;\breve{Y})]
\Big|_{\epsilon=0}
\nonumber\\&
=(1-\lambda)(  H_{\tilde{f}}(V_1,V_2)
- H_{\tilde{f}}(Y,V_1,V_2))
\nonumber\\&
= 0 \,.
\end{align}
We deduce that the perturbed distribution attains the maximal value of the function 
$\tilde{G}_\lambda$. Therefore, we can restrict the union to $|\Vset_1|=|\Vset_2|= |\Xset_1| |\Xset_2|+1$.
 This completes the proof for the cardinality bounds.
\qed

\section{Proof of Theorem~\ref{theo:etMAC_In}}
\label{app:etMAC_In}
Consider the classical MAC $P_{Y|X_1,X_2}$ with entanglement resources between the transmitters (see Figure~\ref{fig:MentangledTx}).

\subsection{Achievability Proof}
We show that for every $\zeta_1,\zeta_2,\eps_0>0$, there exists a $(2^{n(R_1-\zeta_1)},2^{n(R_2-\zeta_2)},n)$ code for $P_{Y|X_1,X_2}$ with entangled transmitters and $\eps_0$-error, provided that $(R_1,R_2)\in \mathcal{R}_{\text{ET}}(P_{Y|X_1,X_2})$. 
To prove achievability, we use coded time sharing
\cite[Sec. 4.5.3]{ElGamalKim:11b}.

Fix a joint distribution $p_{V_0} p_{V_1|V_0} p_{V_2|V_0}$, a bipartite state $\varphi_{A_1 A_2}$, and collection of POVMs $\Lset_k(v_0,v_k)=\{ L_k(x_k|v_0,v_k) \}$ for $k=1,2$. 
Suppose that Alice 1 and Alice 2 share an $n$ copies of the  bipartite state, 
\begin{align}
\varphi_{A_1^n A_2^n}\equiv \varphi_{A_1 A_2}^{\otimes n} \,.
\end{align}
The code construction, encoding with shared entanglement, and decoding procedures are described below.

\vspace{0.2cm}
\subsubsection{Code Construction}
Select a random time-sharing sequence $v_0^n$, according to an i.i.d. distribution, $\prod_{i=1}^n p_{V_0}(v_{0,i})$. Furthermore, select
$2^{nR_1}$ conditionally independent sequences, $v_1^n(m_1)$, $m_1\in [1:2^{nR_1}]$,  each distributed as 
$\prod_{i=1}^n p_{V_1|V_0}\big(v_{1}[i] \,\big| v_{0}[i] \big)$.
In a similar manner,
select
$2^{nR_2}$ sequences,  $v_2^n(m_2)$, %
according to  $\prod_{i=1}^n p_{V_2|V_0}\big(v_{2}[i] \,\big| v_{0}[i] \big)$.

The auxiliary codebooks above are revealed to Alice 1, Alice 2, and Bob.

\vspace{0.2cm}
\subsubsection{Encoder k}
Given the message $m_k\in [1:2^{nR_k}]$ and the codebooks above,  perform the measurement $\bigotimes_{i=1}^n\left(
\Lset_k\big(v_{0}[i],v_{k}[i](m_k) \big) \right)$ on the entangled system $A_k^n$, and transmit the measurement outcome $x_k^n$ through the channel, for $k=1,2$. 

This yields the following input distribution,
\begin{align}
f(x_1^n,x_2^n|m_1,m_2)&=  \trace\left[ \left( L_1^n\big(x_1^n \,\big| v_0^n,v_1^n(m_1) \big) \otimes 
L_2^n \big( x_2^n \,\big| v_0^n,v_2^n(m_2) \big)   \right)  
\varphi_{A_1^n A_2^n}
  \right]
\nonumber\\
&= \prod_{i=1}^n  \trace\left[ \left( L_1\big( x_{1}[i] \,\big| v_{0}[i],v_{1}(m_1)[i] \big) \otimes 
L_2\big( x_{2}[i] \,\big| v_{0}[i],v_{2}(m_2)[i] \big)   \right)  
\varphi_{A_1 A_2}
  \right]	
	\,,
\end{align}
where we use the short notations  $L_k^n\big( x_k^n \,\big| v_0^n,v_k^n \big)\equiv 
\bigotimes_{i=1}^n L_k \big( x_{k}[i] \,\big| v_{0}[i],v_{k}[i] \big)  $, 
for $k=1,2$. %

\vspace{0.2cm}
\subsubsection{Decoder}
Let $\delta>0$ be arbitrarily small.
Find a unique pair $(\hm_1,\hm_2)$ such that $(v_0^n,v_1^n(\hm_1),v_2^n(\hm_2),y^n)\in\Aset_\delta^{(n)}(p_{V_0,V_1, V_2, Y})$, where the marginal distribution $p_{V_0,V_1, V_2, Y}$ 
 is induced by the following joint distribution,
\begin{multline}
p_{V_0,V_1,V_2,X_1,X_2,Y}(v_0,v_1,v_2,x_1,x_2,y)= p_{V_0}(v_0) p_{V_1|V_0}(v_1|v_0) p_{V_2|V_0}(v_2|v_0)\\ \cdot \trace\left[ \left( L_1(x_{1}|v_0,v_{1}) \otimes L_2(x_{2}|v_0,v_{2})   \right) \varphi_{A_1 A_2}
  \right]
		\cdot P_{Y|X_1,X_2}(y|x_1,x_2) \,.
	\label{eq:DirectJointDistribution}
\end{multline}
If there is no such pair $(\hat{m}_1,\hat{m}_2)$, or more than one, declare an error.

\vspace{0.2cm}
\subsubsection{Analysis of Probability of Error}
 We use the notation $\eps_i(\delta)$, $i=1,2,\ldots$,
for terms that tend to zero as $\delta\rightarrow 0$.
At first, suppose that the messages are chosen at random according to a uniform distribution.
By symmetry, we may assume without loss of generality that the transmitters send the messages $M_1=M_2=1$.
Consider the following error events,
\begin{align}
\mathscr{E}_0=& \{  (V_0^n,V_1^n(1),V_2^n(1),Y^n)\notin \Aset_{\delta_1}^{(n)}(p_{V_0,V_1,V_2,Y}) \} \\
\mathscr{E}_1=& \{  (V_0^n,V_1^n(m_1),V_2^n(1),Y^n)\in \Aset_{\delta}^{(n)}(p_{V_0,V_1,V_2,Y})\text{, for some $m_1\neq 1$}  \}\\
\mathscr{E}_2=& \{  (V_0^n,V_1^n(1),V_2^n(m_2),Y^n)\in \Aset_{\delta}^{(n)}(p_{V_0,V_1,V_2,Y})\text{, for some $m_2\neq 1$}  \}\\
\mathscr{E}_3=& \{  (V_0^n,V_1^n(m_1),V_2^n(m_2),Y^n)\in \Aset_{\delta}^{(n)}(p_{V_0,V_1,V_2,Y})\text{, for some $m_1\neq 1$ and $m_2\neq 1$}  \}
\end{align}
 with $\delta_1\equiv \delta/(2 |\Vset_0| |\Vset_1| |\Vset_2|)$.
By the union of events bound, the expected probability of error is bounded by
\begin{align}
\mathbb{E}\left[
P_{e}^{(n)}(\mathscr{C}|1,1)\right] %
&\leq %
 \prob{ \mathscr{E}_0 }%
+ \cprob{ \mathscr{E}_1 }{ \mathscr{E}_0^c }
+ \cprob{ \mathscr{E}_2 }{ \mathscr{E}_0^c }
+ \cprob{ \mathscr{E}_3 }{ \mathscr{E}_0^c } 
\label{eq:PeBsc}
\end{align}
where the expectation on the left-hand side is with respect to the random auxiliary codebooks, and
the conditioning on $M_1=M_2=1$ is omitted from the right-hand side for convenience of notation.
Observe that the (classical) codewords $V_0^n$, $V_1^n(1)$, $V_2^n(1)$,  channel inputs $X_1^n$, $X_2^n$, and channel output $Y^n$, are jointly i.i.d. according to $p_{V_0,V_1,V_2,X_1,X_2,Y}$, as in (\ref{eq:DirectJointDistribution}).
Hence,
by the weak law of large numbers, the first probability term, $\prob{ \mathscr{E}_0 }$, tends to zero as $n\rightarrow\infty$
\cite{CsiszarKorner:82b} \cite[Th. 1.1]{Kramer:08n}.

As for the second error term, we have by the union bound:
\begin{align}
\cprob{ \mathscr{E}_1 }{ \mathscr{E}_0^c }\leq 
\sum_{m_1\neq 1} \cprob{(V_0^n,V_1^n(m_1),V_2^n(1),Y^n)\in \Aset_{\delta}^{(n)}(p_{V_0, V_1,V_2,Y})}{ \mathscr{E}_0^c } \,.
\label{eq:E1bound1}
\end{align}
 Given $\mathscr{E}_0^c$, it follows that $(V_0^n,V_2^n(1))\in \Aset^{\delta}(p_{V_0,V_2}) $. %
Thus, each summand %
is bounded by
\begin{align}
\sum_{(v_0^n,v_2^n)\in \Aset_{\delta}^{(n)}(p_{V_0,V_2})} p_{V_0,V_2}^n(v_0^n,v_2^n) \left[
\sum_{v_1^n,y^n \,:\; (v_0^n,v_1^n,v_2^n,y^n)\in \Aset_{\delta}^{(n)}(p_{V_0,V_1,V_2,Y})}
p_{V_1|V_0}^n(v_1^n|v_0^n)\cdot p_{Y^n|V_0^n,V_2^n}(y^n|v_0^n,v_2^n) \right]
\end{align}
since for every $m_1\neq 1$, the codeword $V_1^n(m_1)$ is conditionally independent of the sequence pair $(V_2^n(1),Y^n)$, given $V_0^n=v_0^n$. 
Then, by standard method-of-types arguments,
the sum within the square brackets is bounded by $2^{-n(I(V_1;Y|V_0 V_2)-\eps_1(\delta))}$ 
\cite{CsiszarKorner:82b} \cite[Th. 1.3]{Kramer:08n}, with %
$
\varepsilon_1(\delta)=
2H(Y|V_0 V_1 V_2)\cdot
\min_{v_0,v_1,v_2}\{p_{V_0,V_1,V_2}(v_0,v_1,v_2) \,:\; p_{V_0,V_1,V_2}(v_0,v_1,v_2) >0 \} 
$. 

Hence, by (\ref{eq:E1bound1}),
\begin{align}
\cprob{ \mathscr{E}_1 }{ \mathscr{E}_0^c }\leq 
2^{-n[I(V_1;Y|V_0 V_2)-R_1-\eps_1(\delta)]} \,.
\label{eq:E1bound2}
\end{align}
Thereby, the term $\cprob{ \mathscr{E}_1 }{ \mathscr{E}_0^c }$  tends to zero as $n\to\infty$, provided that 
\begin{align}
R_1<I(V_1;Y|V_0 V_2)-\eps_1(\delta) \,.
\end{align}
Following similar arguments, the probability terms $ \cprob{ \mathscr{E}_2 }{ \mathscr{E}_0^c }$ and
$ \cprob{ \mathscr{E}_3 }{ \mathscr{E}_0^c } $ also tend to zero, provided that
\begin{align}
R_2<I(V_2;Y|V_0 V_1)-\eps_1(\delta)
\intertext{and}
R_1+R_2<I(V_1 V_2;Y|V_0)-\eps_1(\delta)
 \,.
\end{align}

We conclude that the average probability of error, $\mathbb{E}\left[\overline{P}_e^{(n)}(\code)\right]$, averaged over the messages and the class of random codebooks above, tends to zero as $n\to\infty$ (see (\ref{Equation:Message_Avg_Error})).
Therefore, there must exist a $(2^{nR_1},2^{nR_2},n)$ code such that the message-average error probability, as defined in (\ref{Equation:Message_Avg_Error}), is bounded by 
$\overline{P}_e^{(n)}(\code)< \frac{1}{9}\varepsilon_0$, for a sufficiently large $n$. Based on the arguments in \cite{Cai:14p}, it follows that the rate pair $(R_1,R_2)$ is also achievable with a maximal error criterion, i.e., such that $P_e^{(n)}(\code)\leq \varepsilon_0$.
For completeness, we show this in the appendix.
To show achievability of the regularized formula, apply the coding scheme above to the product channel $P_{Y|X_1,X_2}^{\ell}$.
This completes the achievability proof. \qed

\subsection{Converse Proof}
Consider the classical MAC with entanglement resources between the transmitters. 
We now show the converse part for the regularized characterization. 
 Suppose that Alice 1 and Alice 2 share a bipartite state $\Psi_{E_1 E_2}$. Then, Alice $k$ chooses a message $M_k$ uniformly at random, hence
 the message pair $(M_1,M_2)$ 
 is uniformly distributed over
  $[1:2^{nR_1}]\times [1:2^{nR_2}]$. 
She encodes her message by performing a measurement $\Fset_k^{(M_k)}=\{ F_{x_k^n}^{(M_k)} \}$ on her share of the entangled resources, $E_k$, and sends the measurement outcome $X_k^n$ over the channel. 
 Bob receives the output $Y^n$ and finds an estimate $(\hM_1,\hM_2)=g(Y^n)$ of the message pair.

Now, consider a sequence of codes $(\Psi_n,\Fset_{1n},\Fset_{2n},g_n)$ such that the average probability of error tends to zero, hence
the error probabilities $\prob{ \hM_1\neq M_1 |M_2}$, $\prob{ \hM_2\neq M_2 |M_1}$, and $\prob{ (\hM_1,\hM_2)\neq (M_1,M_2)}$,  are bounded by some
$\alpha_n$ which tends to zero as $n\rightarrow \infty$.
By Fano's inequality \cite{CoverThomas:06b}, it follows that%
\begin{align}
H(M_1|\hM_1,M_2) &\leq n\eps_{1n}\,,\\
H(M_2|\hM_2,M_1) &\leq n\eps_{2n}\,,\\
H(M_1,M_2|\hM_1,\hM_2) &\leq n\eps_{3n}
\label{eq:AFWC2c}
\end{align}
where $\eps_{k\,n}$ tend to zero as $n\rightarrow\infty$.

Hence, 
\begin{align}
nR_1&= H(M_1|M_2)=I(M_1;\hM_1|M_2)+H(M_1|\hM_1 M_2) 
\nonumber\\
&\leq I(M_1;\hM_1|M_2)+n\eps_{1n} \nonumber\\
&\leq I(M_1;Y^n |M_2)+n\eps_{1n}
\label{eq:ConvIneq1}
\end{align}
where the last inequality follows from the data processing inequality.
Following similar arguments, we also have
\begin{align}
nR_2&\leq I(M_2;Y^n |M_1)+n\eps_{2n}
\intertext{and}
n(R_1+R_2)&\leq I(M_1 M_2;Y^n )+n\eps_{3n}
\end{align}
The proof follows by identifying $V_1$ and $V_2$ as $M_1$ and $M_2$, respectively, and taking $V_0=\emptyset$.

\section{Proof of Theorem~\ref{theo:etMAC_Out} (Outer Bound)}
\label{app:etMAC_Out}

We prove the outer bound on the capacity region of the classical MAC with entangled transmitters. Consider the rate bound in (\ref{eq:ConvIneq1}).
Applying the chain rule, we can rewrite this %
as
\begin{align}
n(R_1-\eps_{1n})
&\leq \sum_{i=1}^n I\left(M_1;Y[i] \,\big|\; Y^{i-1}, M_2 \right) \,. %
\label{eq:ConvIneq1a}
\end{align}
where $Y^{i-1}\equiv Y[1],\ldots,Y[i-1]$,
for $i=2,\ldots,n$, and $Y^0\equiv \emptyset$.
Define $V_{0}[i]\equiv Y^{i-1}$, $V_{1}[i]=(M_1,Y^{i-1})$, and $V_{2}[i]=(M_2,Y^{i-1})$, for $i=1,\ldots,n$.
Notice that $V_{1}[i]$ and $V_{2}[i]$ are \emph{not}  conditionally independent given $V_{0}[i]$, because conditioning on the output may introduce statistical dependence between the random messages.

Furthermore, since \mbox{Alice 1} performs a measurement that depends on her message $M_1$ alone, and similarly for \mbox{Alice 2},
the channel inputs $X_{1}[i]$ and $X_2[i]$ can also be obtained as the output of a measurement  as defined below.
Consider a measurement channel
\begin{align}
\widetilde{\mathcal{F}}^{(i,m_1)}_{E_1\to X_1}\otimes 
\widetilde{\mathcal{F}}^{(i,m_2)}_{E_2\to X_2}
\end{align}
such that 
\begin{align}
\widetilde{\mathcal{F}}^{(i,m_1)}_{E_1\to X_1}(Q')&=
\sum_{a_1\in\mathcal{X}_1}
\Bigg[ \sum\limits_{x_1^n\in\Xset_1^n\,:\; x_1[i]=a_1} \trace\left( F_{x_1^n}^{(m_1)}Q' \right) \Bigg]
\ketbra{a_1}
\\
\widetilde{\mathcal{F}}^{(i,m_2)}_{E_2\to X_2}(Q'')&=
\sum_{a_2\in\mathcal{X}_2}
\Bigg[ \sum\limits_{x_2^n\in\Xset_2^n\,:\; x_2[i]=a_2} \trace\left( F_{x_2^n}^{(m_2)}Q'' \right) \Bigg]
\ketbra{a_2}
\end{align}
for every pair of operators $Q'$ and $Q''$ on $\Hset_{E_1}$
and $\Hset_{E_2}$, respectively. 
Let $\Psi_{E_1 E_2}=\sum_{j,\ell} Q'_j\otimes Q''_\ell$ be an arbitrary decomposition of the bipartite state. Hence, by linearity,
\begin{align}
&\left(
\widetilde{\mathcal{F}}^{(i,m_1)}_{E_1\to X_1}\otimes 
\widetilde{\mathcal{F}}^{(i,m_2)}_{E_2\to X_2}\right)(\Psi_{E_1 E_2})
\nonumber\\
&=\sum_{t,r} \left(
\widetilde{\mathcal{F}}^{(i,m_1)}_{E_1\to X_1}\otimes 
\widetilde{\mathcal{F}}^{(i,m_2)}_{E_2\to X_2}\right)(Q'_j\otimes Q''_\ell)
\nonumber\\
&=
\sum_{j,\ell}
\sum_{a_1\in\mathcal{X}_1}
\Bigg[ \sum\limits_{x_1^n\in\Xset_1^n\,:\; x_1[i]=a_1} \trace\left( F_{x_1^n}^{(m_1)}Q_j' \right) \Bigg]
\ketbra{a_1}\otimes 
\sum_{a_2\in\mathcal{X}_2}
\Bigg[ \sum\limits_{x_2^n\in\Xset_2^n\,:\; x_2[i]=a_2} \trace\left( F_{x_2^n}^{(m_2)}Q_\ell'' \right) \Bigg] \ketbra{a_2}
\nonumber\\
&=
\sum_{(a_1,a_2)\in\mathcal{X}_1\times \mathcal{X}_2}
\left[ \sum_{\substack{ (x_1^n,x_2^n)\in\Xset_1^n\times \Xset_2^n\,:\; \\ x_1[i]=a_1\,,\;
 x_2[i]=a_2}}\; \sum_{j,\ell} \trace\left( (F_{x_1^n}^{(m_1)}\otimes F_{x_2^n}^{(m_2)})(Q_j'\otimes Q_\ell'') \right) \right]
\ketbra{a_1,a_2} 
\nonumber\\
&=
\sum_{(a_1,a_2)\in\mathcal{X}_1\times \mathcal{X}_2}
\left[ \sum_{\substack{ (x_1^n,x_2^n)\in\Xset_1^n\times \Xset_2^n:\; \\ x_1[i]=a_1\,,\;
 x_2[i]=a_2}} \trace\left( (F_{x_1^n}^{(m_1)}\otimes F_{x_2^n}^{(m_2)})\Psi_{E_1 E_2} \right) \right]
\ketbra{a_1,a_2} 
\label{eq:Converse_Measurement}
\end{align}
as expected.
We deduce that the channel inputs $X_{1}[i]$ and $X_2[i]$ can be obtained from a product of  measurements of the form $L_1(x_{1}|i,V_{0}[i],V_{1}[i])$ and $L_2(x_{2}|i,V_{0}[i],V_{2}[i])$, as required.

 Then, we can rewrite (\ref{eq:ConvIneq1a}) as
\begin{align}
R_1-\eps_{1n}
&\leq \frac{1}{n}\sum_{i=1}^n I\left(V_{1}[i];Y[i] \,\big|\; V_{0}[i], V_{2}[i] \right) \nonumber\\
&=  I\left(V_{1}[J];Y[J] \,\big|\; V_{0}[J], V_{2}[J], J \right) 
\label{eq:ConvIneq1b}
\end{align}
where the index $J$ is drawn  uniformly at random from $\{1,\ldots,n\}$, and it is uncorrelated with the previous systems.
Following the same considerations,
\begin{align}
R_2-\eps_{2n}
&\leq  I\left(V_{2}[J];Y[J] \,\big|\; V_{0}[J], V_{1} [J], J \right) 
\label{eq:ConvIneq2}
\intertext{and}
R_1+R_2-\eps_{3n}
&\leq  I\left(V_{1}[J] V_{2}[J];Y[J] \,\big|\; V_{0}[J], J \right) \,. 
\label{eq:ConvIneq3}
\end{align}
The proof follows by defining $V_0\equiv (J,V_{0}[J])$, $V_k\equiv V_{k}[J] $, for $k=1,2$, then
$X_k\equiv X_{k}[J]$, and $Y\equiv Y[J]$. 
\qed

\section{Proof of Theorem~\ref{theo:etMAC_Conferencing} (Classical Conferencing)}
\label{app:etMAC_Conferencing}
Consider the classical MAC $P_{Y|X_1,X_2}$ with entanglement resources and classical conferencing between the transmitters (see Figure~\ref{fig:MentangledTx_Conferencing}).

\subsection{Achievability Proof}
We show that for every $\delta_1,\delta_2,\eps_0>0$, there exists  $(2^{n(R_1-\delta_1)},2^{n(R_2-\delta_2)},n)$ code that has an  $\eps_0$-error, provided that $(R_1,R_2)\in \mathcal{R}_{\text{ET-C}}(P_{Y|X_1,X_2},C_{12},C_{21})$. 
We use the classical rate splitting method due to Willems \cite{Willems:83p}.

As in the previous section,
fix a joint distribution $p_{V_0} p_{V_1|V_0} p_{V_2|V_0}$, a bipartite state $\varphi_{A_1 A_2}$, and collection of POVMs $\Lset_k(v_0,v_k)%
$ for $k=1,2$. 
Suppose that Alice 1 and Alice 2 share %
$%
\varphi_{A_1^n A_2^n}\equiv \varphi_{A_1 A_2}^{\otimes n} %
$. %

The code construction, conferencing protocol, encoding with shared entanglement, and decoding procedures are given below.

\vspace{0.2cm}
\subsubsection{Code Construction}
Split the rates of each user into $R_k=R_{k}'+R_{k}''$ for 
$k=1,2$.
We think of $R_{k}'$ as the rate of \emph{common} information that is shared between the transmitters through conferencing, and $R_{k}''$ as the rate of the \emph{private} information of each transmitter.

Select $2^{n(R_{1}'+R_{2}')}$ 
sequences 
$v_0^n(m_{1}',m_{2}')$, 
for $m_{k}'\in [1:2^{nR_{k}'}]$, independently at random, each
 i.i.d.  $\sim p_{V_0}$. For every $(m_{1}',m_{2}')%
$,
select
$2^{nR_{1}''}$ conditionally independent sequences, $v_1^n(m_{1}',m_{2}',m_{1}'')$, $m_{1}''\in [1:2^{nR_{1}''}]$,  each distributed as 
$\prod_{i=1}^n p_{V_1|V_0}\big(v_{1}[i] \,\big| v_{0}(m_{1}',m_{2}')[i] \big)$.
In a similar manner,
select
$2^{nR_{2}''}$ sequences,  $v_2^n(m_{1}',m_{2}',m_{2}'')$, %
according to  $\prod_{i=1}^n p_{V_2|V_0}\big(v_{2}[i] \,\big| v_{0}(m_1',m_2')[i] \big)$.
The auxiliary codebooks are revealed to all parties. %

\vspace{0.2cm}
\subsubsection{Conferencing}
The encoders  share the messages $m_1'$ and $m_2'$ between them through a single round of conferencing ($T=1$). This is permissible if the common information rates are each bounded by the respective conferencing capacity, i.e.,
\begin{align}
R_1' \leq C_{12} \,,\;
R_2' \leq C_{21} %
\label{eq:Permissible_Rates}
\end{align}
as this satisfies the conferencing limits in (\ref{eq:Permissible}).

\vspace{0.2cm}
\subsubsection{Encoder 1}
Given the message $(m_1',m_1'')$, the conferencing message $m_2'$, and the codebooks above,  perform the measurement $\bigotimes_{i=1}^n\left(
\Lset_1\big(v_{0}(m_1',m_2')[i],v_{1}(m_1',m_2',m_1'')[i] \big) \right)$ on the entangled system $A_1^n$, and transmit the measurement outcome $x_1^n$ through the channel. 

\vspace{0.2cm}
\subsubsection{Encoder 2}
Given the message $(m_2',m_2'')$, the conferencing message $m_1'$, and the codebooks above,  perform the measurement $\bigotimes_{i=1}^n\left(
\Lset_2\big(v_{0}(m_1',m_2')[i],v_{2}(m_1',m_2',m_2'')[i] \big) \right)$ on the entangled system $A_2^n$, and transmit the measurement outcome $x_2^n$ through the channel.

The resulting input distribution is given by
\begin{align}
f(x_1^n,x_2^n|m_1',m_1'',m_2',m_2'')
&= \prod_{i=1}^n  \trace\left[ \left( L_1\big( x_{1}[i] \,\big| v_{0}[i],v_{1}[i] \big) \otimes 
L_2\big( x_{2}[i] \,\big| v_{0}[i],v_{2}[i] \big)   \right)  
\varphi_{A_1 A_2}
  \right]	
\end{align}
where $v_0^n\equiv v_{0}^n(m_1',m_2')$ and $v_k^n\equiv v_{k}^n(m_1',m_2',m_k'')$ for $k=1,2$.

\vspace{0.2cm}
\subsubsection{Decoder}
In a similar manner as in the previous proof,
find a unique tuple $(\hm_1',\hm_1'',\hm_2',\hm_2'')$ such that $(v_0^n(\hm_1',\hm_2'),$ $v_1^n(\hm_1',\hm_2',\hm_1''),$ $v_2^n(m_1',m_2',\hm_2''),$ $y^n)\in\Aset_\delta^{(n)}(p_{V_0,V_1, V_2, Y})$. 
As before, the joint distribution is %
\begin{multline}
p_{V_0,V_1,V_2,X_1,X_2,Y}(v_0,v_1,v_2,x_1,x_2,y)= p_{V_0}(v_0) p_{V_1|V_0}(v_1|v_0) p_{V_2|V_0}(v_2|v_0)\\ \cdot \trace\left[ \left( L_1(x_{1}|v_0,v_{1}) \otimes L_2(x_{2}|v_0,v_{2})   \right) \varphi_{A_1 A_2}
  \right]
		\cdot P_{Y|X_1,X_2}(y|x_1,x_2) \,.
	\label{eq:DirectJointDistribution_Conferencing}
\end{multline}
If there is none, or more than one such pair, declare an error.

\subsubsection{Analysis of Probability of Error}
At first, suppose that the messages are uniformly distributed.
By symmetry, we may assume without loss of generality that the transmitters send  $M_1'=M_1''=M_2'=M_2''=1$.

We consider the error events below, whereby the correct messages do not satisfy the decoding rule (the event $\mathscr{E}_0$ below), or alternatively, the decoder has ambiguity for the common information alone ($\mathscr{E}_0'$), for one of the private messages ($\mathscr{E}_1$ and $\mathscr{E}_2$), both private messages
($\mathscr{E}_3$), the common information and one private message ($\mathscr{E}_4$ and $\mathscr{E}_5$), or all messages ($\mathscr{E}_6$):
\begin{align}
\mathscr{E}_0=& \{  (V_0^n(1,1),
V_1^n(1,1,1),V_2^n(1,1,1),Y^n)\notin \Aset_{\delta_1}^{(n)}(p_{V_0,V_1,V_2,Y}) \} \\
\mathscr{E}_0'=& \{  (V_0^n(m_1',m_2'),V_1^n(m_1',m_2',1),V_2^n(m_1',m_2',1),Y^n)\in \Aset_{\delta}^{(n)}(p_{V_0,V_1,V_2,Y})\text{, for some $(m_1',m_2')\neq (1,1)$}  \}\\
\mathscr{E}_1=& \{  (V_0^n(1,1),V_1^n(1,1,m_1''),V_2^n(1,1,1),Y^n)\in \Aset_{\delta}^{(n)}(p_{V_0,V_1,V_2,Y})\text{, for some $m_1''\neq 1$}  \}\\
\mathscr{E}_2=& \{  (V_0^n(1,1),V_1^n(1,1,1),V_2^n(1,1,m_2''),Y^n)\in \Aset_{\delta}^{(n)}(p_{V_0,V_1,V_2,Y})\text{, for some $m_2''\neq 1$}  \}\\
\mathscr{E}_3=& \{  (V_0^n(1,1),V_1^n(1,1,m_1''),V_2^n(1,1,m_2''),Y^n)\in \Aset_{\delta}^{(n)}(p_{V_0,V_1,V_2,Y})\text{, for some $m_1''\neq 1$ and $m_2''\neq 1$}  \}\\
\mathscr{E}_4=& \{  (V_0^n(m_1',m_2'),V_1^n(m_1',m_2',m_1''),V_2^n(m_1',m_2',1),Y^n)\in \Aset_{\delta}^{(n)}(p_{V_0,V_1,V_2,Y})\text{, for some $(m_1',m_2')\neq (1,1)$ and $m_1''\neq 1$}  \}\\
\mathscr{E}_5=& \{  (V_0^n(m_1',m_2'),V_1^n(m_1',m_2',1),V_2^n(m_1',m_2',m_2''),Y^n)\in \Aset_{\delta}^{(n)}(p_{V_0,V_1,V_2,Y})\text{, for some $(m_1',m_2')\neq (1,1)$ and $m_2''\neq 1$}  \}\\
\mathscr{E}_6=& \{  (V_0^n(m_1',m_2'),V_1^n(m_1',m_2',m_1''),V_2^n(m_1',m_2',m_2''),Y^n)\in \Aset_{\delta}^{(n)}(p_{V_0,V_1,V_2,Y})\text{, for some $(m_1',m_2')\neq (1,1)$, $m_1''\neq 1$, }\nonumber\\& \text{and $m_2''\neq 1$}  \}
\end{align}
 with $\delta_1\equiv \delta/(2 |\Vset_0| |\Vset_1| |\Vset_2|)$.
By the union of events bound, the expected probability of error is bounded by
\begin{align}
\mathbb{E}\left[
P_{e}^{(n)}(\mathscr{C}|1,1)\right] %
&\leq %
 \prob{ \mathscr{E}_0 }+\prob{ \mathscr{E}_0' }+\cprob{ \mathscr{E}_1 }{\mathscr{E}_0^c}
\nonumber\\&
+ \cprob{ \mathscr{E}_2 }{ \mathscr{E}_0^c }
+ \cprob{ \mathscr{E}_3 }{ \mathscr{E}_0^c }
+ \prob{ \mathscr{E}_4 } 
\label{eq:PeBsc_Conferencing}
\end{align}
where the expectation %
is with respect to the random auxiliary codebooks, and
the conditioning on $M_1'=M_1''=M_2'=M_2''=1$ is omitted from the right-hand side for convenience of notation.
Observe that the  codewords $V_0^n(1,1)$, $V_1^n(1,1,1)$, $V_2^n(1,1,1)$,  channel inputs $X_1^n$, $X_2^n$, and channel output $Y^n$, are jointly i.i.d. according to $p_{V_0,V_1,V_2,X_1,X_2,Y}$, as in (\ref{eq:DirectJointDistribution_Conferencing}).
As in the proof of Theorem~\ref{theo:etMAC_In}, we observe that $\prob{ \mathscr{E}_0 }$, tends to zero as $n\rightarrow\infty$
by the weak law of large numbers. %

Furthermore,
by the union bound, the second error term is bounded by 
\begin{align}
\prob{ \mathscr{E}_0' }\leq 
\sum_{(m_1',m_2')\neq (1,1)} \prob{(V_0^n(m_1',m_2'),V_1^n(m_1',m_2',1),V_2^n(m_1',m_2',1),Y^n)\in \Aset_{\delta}^{(n)}(p_{V_1,V_2,Y})} \,.
\label{eq:E0pbound1_Conferencing}
\end{align}
By standard method-of-types arguments,
each term is then bounded by $2^{-n(I(V_0 V_1 V_2;Y)-\eps_0(\delta))}$ 
\cite{CsiszarKorner:82b} \cite[Th. 1.3]{Kramer:08n}, with %
$
\varepsilon_0(\delta)=
2H(Y)\cdot
\min_{v_0,v_1,v_2}\{p_{V_0,V_1,V_2}(v_0,v_1,v_2) \,:\; p_{V_0,V_1,V_2}(v_0,v_1,v_2) >0 \} 
$. 
Since $V_0 \Cbar (V_1,V_2)\Cbar Y$ form a Markov chain, we have $I(V_0 V_1 V_2;Y)=I(V_1 V_2;Y)$.
It follows that 
\begin{align}
\prob{ \mathscr{E}_0' }\leq 
2^{-n[I(V_1 V_2;Y)-R_1'-R_2'-\eps_0(\delta)]} \,.
\label{eq:E0bound2_Conferencing}
\end{align}
Thereby, the term $\prob{ \mathscr{E}_0' }$  tends to zero as $n\to\infty$, provided that 
\begin{align}
R_1'+R_2'<I(V_1 V_2;Y)-\eps_0(\delta) \,.
\end{align}

Now, consider the third error term:
\begin{align}
\cprob{ \mathscr{E}_1 }{ \mathscr{E}_0^c } \leq 
\sum_{m_1''\neq 1} \prob{(V_0^n(1,1),V_1^n(1,1,m_1''),V_2^n(1,1,1),Y^n)\in \Aset_{\delta}^{(n)}(p_{V_0, V_1,V_2,Y})} \,.
\label{eq:E1bound2_Conferencing}
\end{align}
 Given $\mathscr{E}_0^c$, it follows that $(V_0^n(1,1),V_2^n(1,1,1))\in \Aset^{\delta}(p_{V_0 V_2}) $. %
Thus, each summand %
is bounded by
\begin{align}
\sum_{(v_0^n,v_2^n)\in \Aset_{\delta}^{(n)}(p_{V_0,V_2})} p_{V_0,V_2}^n(v_0^n,v_2^n) \left[
\sum_{v_1^n,y^n \,:\; (v_0^n,v_1^n,v_2^n,y^n)\in \Aset_{\delta}^{(n)}(p_{V_0,V_1,V_2,Y})}
p_{V_1|V_0}^n(v_1^n|v_0^n)\cdot p_{Y^n|V_0^n,V_2^n}(y^n|v_0^n,v_2^n) \right]
\end{align}
since for every $m_1''\neq 1$, the codeword $V_1^n(1,1,m_1'')$ is conditionally independent of the sequence pair $(V_2^n(1,1,1),Y^n)$, given $V_0^n(1,1)=v_0^n$. 
As the sum within the square brackets is bounded by $2^{-n(I(V_1;Y|V_0 V_2)-\eps_1(\delta))}$, for %
$
\varepsilon_1(\delta)=
2H(Y|V_0 V_1 V_2)\cdot
\min_{v_0,v_1,v_2}\{p_{V_0,V_1,V_2}(v_0,v_1,v_2) \,:\; p_{V_0,V_1,V_2}(v_0,v_1,v_2) >0 \} 
$ 
(see \cite{CsiszarKorner:82b} \cite[Th. 1.3]{Kramer:08n}), we have 
\begin{align}
\cprob{ \mathscr{E}_1 }{ \mathscr{E}_0^c }\leq 
2^{-n[I(V_1;Y|V_0 V_2)-R_1''-\eps_1(\delta)]} \,.
\label{eq:E1bound3_Conferencing}
\end{align}
Thereby, the term $\cprob{ \mathscr{E}_1 }{ \mathscr{E}_0^c }$  tends to zero as $n\to\infty$, provided that 
$%
R_1''<I(V_1;Y|V_0 V_2)-\eps_1(\delta) %
$. %
Since $R_1=R_1'+R_1''$, this requires
\begin{align}
R_1<I(V_1;Y|V_0 V_2)+R_1'-\eps_1(\delta)  \,.
\end{align}
Recall that by (\ref{eq:Permissible_Rates}), $R_{1}'\leq C_{12}$.

Following similar arguments, the probability terms $ \cprob{ \mathscr{E}_2 }{ \mathscr{E}_0^c }$ and $ \cprob{ \mathscr{E}_3 }{ \mathscr{E}_0^c }$  also tend to zero, provided that
\begin{align}
R_2&<I(V_2;Y| V_0 V_1)+R_2'-\eps_1(\delta)
\intertext{and}
R_1+R_2&<I(V_1 V_2;Y|V_0)+R_1'+R_2'-\eps_1(\delta)
 \,.
\end{align}
By (\ref{eq:Permissible_Rates}), $R_{1}'\leq C_{12}$ and $R_{2}'\leq C_{21}$.
Similarly, the remaining error terms, $\cprob{ \mathscr{E}_j'' }{ \mathscr{E}_0^c }$, $j\in\{4,5,6\}$, tend to zero if 
$R_1'+R_2'+R_1''<I(V_1 V_2;Y)-\eps_1(\delta)$,
$R_1'+R_2'+R_2''<I(V_1 V_2;Y)-\eps_1(\delta)$, and
$R_1'+R_2'+R_1''+R_2''<I(V_1 V_2;Y)-\eps_1(\delta)$.
Thus, it suffices to require
\begin{align}
R_1+R_2&<I(V_1 V_2;Y)-\eps_1(\delta)
\,.
\end{align}

We conclude that the average probability of error, $\mathbb{E}\left[\overline{P}_e^{(n)}(\code)\right]$, averaged over the messages and the class of random codebooks above, tends to zero as $n\to\infty$ (see (\ref{Equation:Message_Avg_Error})).
Therefore, there must exist a $(2^{nR_1},2^{nR_2},n)$ code such that the message-average error probability vanishes. %
Based on the arguments in the appendix, it follows that the rate pair $(R_1,R_2)$ is also achievable with a maximal error criterion, i.e., such that $P_e^{(n)}(\code)\leq \varepsilon_0$. Achievability for the regularized formula follows by applying the coding scheme above to the product MAC $P_{Y|X_1,X_2}^{\ell}$.
This completes the achievability proof.

\subsection{Converse Proof}
Consider the classical MAC with entanglement resources and classical conferencing between the transmitters. 
We now show the converse part. 
 Suppose that Alice 1 and Alice 2 share a bipartite state $\Psi_{E_1 E_2}$. Then, Alice $k$ chooses a message $M_k$, uniformly at random from $[1:2^{nR_k}]$, for $k=1,2$. They perform a conferencing protocol, whereby Alice 1 sends 
 $\omega_{1\to 2}^T=s_{1\to 2}^T(M_1,M_2)$ to Alice 2, and Alice 2 sends 
  $\omega_{2\to 1}^T=s_{2\to 1}^T(M_1,M_2)$ to Alice 1.
Next, Alice 1 encodes her message $M_1$ by performing a measurement $\Fset_1^{(M_1,\omega_{2\to 1}^T)}=\{ F_{x_1^n}^{(M_1,\omega_{2\to 1}^T)} \}$ on her share of the entangled resources, $E_1$, and transmits the measurement outcome $X_1^n$ over the channel. 
In the same manner, Alice 2 performs an encoding measurement $\Fset_2^{(M_2,\omega_{1\to 2}^T)}$ that produces her transmission, $X_2^n$. 
 Bob receives the output $Y^n$ and finds an estimate $(\hM_1,\hM_2)=g(Y^n)$ of the message pair.

Now, consider a sequence of
$(C_{12},C_{21})$-permissible
codes $(\Psi_n,s_{1\to 2(n)},s_{2\to 1(n)},\Fset_{1(n)},\Fset_{2(n)},g_n)$ such that the average probability of error tends to zero, hence
the error probabilities $\prob{ \hM_1\neq M_1 |M_2}$, $\prob{ \hM_2\neq M_2 |M_1}$, and $\prob{ (\hM_1,\hM_2)\neq (M_1,M_2)}$,  are bounded by some
$\alpha_n$ which tends to zero as $n\rightarrow \infty$.
By Fano's inequality \cite{CoverThomas:06b}, 
\begin{align}
H(M_1|\hM_1,M_2) &\leq n\eps_{1n}\,,\\
H(M_2|\hM_2,M_1) &\leq n\eps_{2n}\,,\\
H(M_1,M_2|\hM_1,\hM_2) &\leq n\eps_{3n}
\label{eq:AFWC2c_Conferencing}
\end{align}
where $\eps_{k\,n}$ tend to zero as $n\rightarrow\infty$.
Since conditioning cannot increase entropy, we also have
\begin{align}
H(M_1|\hM_1,M_2,\omega_{1\to 2}^T,\omega_{2\to 1}^T) &\leq n\eps_{1n}\,,\\
H(M_2|\hM_2,M_1,\omega_{1\to 2}^T,\omega_{2\to 1}^T) &\leq n\eps_{2n}\,,\\
H(M_1,M_2|\hM_1,\hM_2,\omega_{1\to 2}^T,\omega_{2\to 1}^T) &\leq n\eps_{3n} \,.
\label{eq:AFWC2c_Conferencing_s}
\end{align}

Now, 
\begin{align}
nR_1&= H(M_1|M_2)=I(M_1;\hM_1  \omega_{1\to 2}^T,\omega_{2\to 1}^T|M_2)+H(M_1|\hM_1 M_2 \omega_{1\to 2}^T \omega_{2\to 1}^T) 
\nonumber\\
&\leq I(M_1;\hM_1 \omega_{1\to 2}^T \omega_{2\to 1}^T|M_2)+n\eps_{1n} \nonumber\\
&\leq I(M_1;Y^n \omega_{1\to 2}^T \omega_{2\to 1}^T |M_2)+n\eps_{1n}
\label{eq:ConvIneq1SC_Conferencing}
\end{align}
where the last inequality follows from the data processing inequality. Applying the chain rule, we can rewrite this %
as
\begin{align}
n(R_1-\eps_{1n})
&\leq I(M_1;  \omega_{1\to 2}^T  \omega_{2\to 1}^T|M_2)+ I(M_1;Y^n |M_2, \omega_{1\to 2}^T \omega_{2\to 1}^T)
\nonumber\\
&= \sum_{t=1}^T I\left(M_1;  \omega_{1\to 2}[t] \omega_{2\to 1}[t] \,\big|\; M_2 , \omega_{1\to 2}^{t-1}\omega_{2\to 1}^{t-1}\right)+ I(M_1;Y^n |M_2, \omega_{1\to 2}^T \omega_{2\to 1}^T)
\label{eq:ConvIneq1_Conferencing}
\end{align}
Since $\omega_{2\to 1}[t]$ is a deterministic function of $(M_2 , \omega_{1\to 2}^{t-1})$, this reduces to 
\begin{align}
n(R_1-\eps_{1n})
&\leq \sum_{t=1}^T I\left(M_1;  \omega_{1\to 2}[t]  \,\big|\; M_2 , \omega_{1\to 2}^{t-1}\omega_{2\to 1}^{t-1}\right)+ \sum_{i=1}^n I(M_1;Y^n |M_2, \omega_{1\to 2}^T \omega_{2\to 1}^T)
\nonumber\\
&\leq nC_{12}+  \sum_{i=1}^n I(M_1;Y^n |M_2, \omega_{1\to 2}^T \omega_{2\to 1}^T)
\label{eq:ConvIneq1_Conferencing2}
\end{align}
as $\sum_{t=1}^T\log|\Omega_{12}^{(t)}|\leq nC_{12}$ for a $(C_{12},C_{21})$-permissible code (see (\ref{eq:Permissible})).
The converse part for the regularized capacity formula follows by identifying $V_0$, $V_1$, and $V_2$ with $\omega_{1\to 2}^T \omega_{2\to 1}^T$, $M_1$, and $M_2$. Then, by \cite[Eq. (20)]{Willems:83p}, $V_1$ and $V_2$ are statistically independent given $V_0$.

As for the single-letter outer bound, we now have
\begin{align}
n(R_1-\eps_{1n})
&\leq nC_{12}+ \sum_{i=1}^n I\left(M_1;Y[i] \,\big|\; Y^{i-1},\omega_{1\to 2}^T, \omega_{1\to 2}^T\omega_{2\to 1}^T, M_2 \right)  %
\,.
\label{eq:ConvIneq1_Conferencing_1}
\end{align}
Then,
define $V_{0}[i]\equiv (Y^{i-1},\omega_{1\to 2}^T \omega_{2\to 1}^T)$, $V_{1}[i]=(M_1,Y^{i-1},\omega_{1\to 2}^T \omega_{2\to 1}^T)$, and $V_{2}[i]=(M_2,Y^{i-1},\omega_{1\to 2}^T \omega_{2\to 1}^T)$, for $i=1,\ldots,n$.

Furthermore, since \mbox{Alice 1} performs a measurement that depends on her message $m_1$ and 
$\omega_{2\to 1}^T$ alone, and similarly for \mbox{Alice 2},
the channel inputs $X_{1}[i]$ and $X_2[i]$ can also be obtained as the output of a measurement  as defined below.
Consider a measurement channel
\begin{align}
\widetilde{\mathcal{F}}^{(i,m_1)}_{E_1\to X_1}\otimes 
\widetilde{\mathcal{F}}^{(i,m_2)}_{E_2\to X_2}
\end{align}
such that 
\begin{align}
\widetilde{\mathcal{F}}^{(i,m_1, \omega_{2\to 1}^T)}_{E_1\to X_1}(Q')&=
\sum_{a_1\in\mathcal{X}_1}
\Bigg[ \sum\limits_{x_1^n\in\Xset_1^n\,:\; x_1[i]=a_1} \trace\left( F_{x_1^n}^{(m_1, \omega_{2\to 1}^T)}Q' \right) \Bigg]
\ketbra{a_1}
\\
\widetilde{\mathcal{F}}^{(i,m_2,\omega_{1\to 2}^T )}_{E_2\to X_2}(Q'')&=
\sum_{a_2\in\mathcal{X}_2}
\Bigg[ \sum\limits_{x_2^n\in\Xset_2^n\,:\; x_2[i]=a_2} \trace\left( F_{x_2^n}^{(m_2,\omega_{1\to 2}^T )}Q'' \right) \Bigg]
\ketbra{a_2}
\end{align}
for every pair of operators $Q'$ and $Q''$ on $\Hset_{E_1}$
and $\Hset_{E_2}$, respectively. 
Let $\Psi_{E_1 E_2}=\sum_{j,\ell} Q'_j\otimes Q''_\ell$ be an arbitrary decomposition of the bipartite state. Hence, following the same steps as in (\ref{eq:Converse_Measurement}), we have %
\begin{align}
&\left(
\widetilde{\mathcal{F}}^{(i,m_1)}_{E_1\to X_1}\otimes 
\widetilde{\mathcal{F}}^{(i,m_2)}_{E_2\to X_2}\right)(\Psi_{E_1 E_2})
=
\nonumber\\&
\sum_{(a_1,a_2)\in\mathcal{X}_1\times \mathcal{X}_2}
\left[ \sum_{\substack{ (x_1^n,x_2^n)\in\Xset_1^n\times \Xset_2^n:\; \\ x_1[i]=a_1\,,\;
 x_2[i]=a_2}} \trace\left( (F_{x_1^n}^{(m_1, \omega_{2\to 1}^T)}\otimes F_{x_2^n}^{(m_2, \omega_{1\to 2}^T)})\Psi_{E_1 E_2} \right) \right]
\ketbra{a_1,a_2} 
\end{align}
as expected.
We deduce that the channel inputs $X_{1}[i]$ and $X_2[i]$ can be obtained from a product of  measurements of the form $L_1(x_{1}|i,V_{0}[i],V_{1}[i])$ and $L_2(x_{2}|i,V_{0}[i],V_{2}[i])$, as required.

 Then, we can rewrite (\ref{eq:ConvIneq1_Conferencing_1}) as
\begin{align}
R_1-\eps_{1n}
&\leq \frac{1}{n}\sum_{i=1}^n I\left(V_{1}[i];Y[i] \,\big|\; V_{0}[i], V_{2}[i] \right)+C_{12} \nonumber\\
&=  I\left(V_{1}[J];Y[J] \,\big|\; V_{0}[J], V_{2}[J], J |V_0[J],J \right) +C_{12}
\end{align}
where the index $J$ is drawn  uniformly at random from $\{1,\ldots,n\}$, and it is uncorrelated with the previous systems.
Following the same considerations,
\begin{align}
R_2-\eps_{2n}
&\leq  I\left(V_{2}[J];Y[J] \,\big|\; V_{0}[J], V_{1} [J], J \right) +C_{21}
\intertext{and}
R_1+R_2-\eps_{3n}
&\leq  I\left(V_{1}[J] V_{2}[J];Y[J] \,\big|\; V_{0}[J], J \right)+C_{12}+C_{21} \,. 
\end{align}

In addition, by (\ref{eq:AFWC2c_Conferencing}),
\begin{align}
R_1+R_2-\eps_{3n}%
&\leq \frac{1}{n} I(M_1 M_2;\hM_1 \hM_2) \nonumber\\
&\leq \frac{1}{n}I(M_1 M_2;Y^n) \nonumber\\
&=  \frac{1}{n} \sum_{i=1}^n I(M_1 M_2;Y[i] \,|\; Y^{i-1})\nonumber\\
&\leq   I(V_1[J] V_2[J] J ;Y[J] )
\label{eq:ConvIneqSum_Conferencing}
\end{align}
The proof follows by defining  $V_k\equiv (J,V_{k}[J]) $, for $k=0,1,2$, then
$X_k\equiv X_{k}[J]$, and $Y\equiv Y[J]$. 
\qed

\section{Summary and Discussion
}
\label{Section:Discussion}

\subsection{Summary}
To summarize, 
we have considered
communication over a two-user classical multiple-access channel
(MAC) $P_{Y|X_1,X_2}$ with  entanglement resources shared
between the transmitters a priori, before communication begins, as illustrated in Figure~\ref{fig:MentangledTx}.
Our main results are inner and outer bounds, as well as a regularized  characterization, for the capacity region of the \emph{general} MAC with entangled transmitters.
We have also  shown that the 
examples by Leditzky et al.  \cite{LeditzkyAlhejjiLevinSmith:20p} are a special case and follow from our main result in 
Theorem~\ref{theo:etMAC_In}.

In addition, we  observe the following change of behavior.
In general, achievable communication rates may also depend on the error criterion.
For a classical MAC
without entanglement resources, %
it has long been known that
the relaxation to a message-average error criterion can lead to strictly higher achievable rates, when compared with a maximal error criterion \cite{Dueck:78p} (see Remark~\ref{Remark:Classical_Maximal_Error}).
Here, however,  the capacity region with entangled transmitters remains the same, whether we consider a message-average or a maximal error criterion.

 The auxiliary systems $A_1$ and $A_2$ in our characterization have unbounded dimensions, as mentioned in Remarks \ref{Remark:Dimension}-\ref{rem:Uncomput}. %
Although one can always compute an achievable region by simply choosing the dimension of $A_1$ and $A_2$, the optimal rates cannot be 
 computed exactly in general. 
 The lack of a dimension bound for the reference system plagues many other 
 quantum models in network information theory, such as 
the MAC with entangled transmitter and receiver \cite{HsiehDevetakWinter:08p},
the  broadcast channel (see Discussion section in \cite{DupuisHaydenLi:10p}), wiretap channel \cite[Remark 5]{QiSharmaWilde:18p},
and squashed entanglement \cite[Section 1]{LiWinter:14p}.
Here, the problem is inherent to the communication scenario.
The examples due to 
Leditzky et al.  
\cite{LeditzkyAlhejjiLevinSmith:20p}
demonstrate that for some channels,
the capacity region can be achieved with a finite entanglement rate (see Example~\ref{Example:Magic_Square}), while for other channels,
achieving the full capacity region requires an infinite amount of entanglement resources (see Example~\ref{Example:Slofstra_Vidick}).

\subsection{Unbounded Dimensions}
In the resource theory literature,
the set of conditional distributions
$p_{X_1,X_2|V_1,V_2}$ arising from all choices of finite dimensional Hilbert space $\mathcal{H}_{A_1}$ and $\mathcal{H}_{A_2}$, all measurements $\mathcal{L}_1(v_1)\otimes \mathcal{L}_2(v_2)$, and all $\ket{\phi_{A_1 A_2}}$, is called the family of quantum spatial correlation matrices, and it is denoted by \cite{DykemaPaulsen:16p}
\begin{align*}
C_{ q s}(|\mathcal{X}_1|,|\mathcal{X}_2|)
\,.
\end{align*}
Furthermore, the quantum commuting family,
$C_{qc}(|\mathcal{X}_1|,|\mathcal{X}_2|)$,
is the set of commuting quantum correlation matrices,
i.e.,  such that the measurement operators for Alice 1 and Alice 2 commute, and the quantum approximate family $C_{qa}(|\mathcal{X}_1|,|\mathcal{X}_2|)$ is the closure of the quantum spatial family, $C_{qa}(|\mathcal{X}_1|,|\mathcal{X}_2|)\equiv
\overline{C_{qs}(|\mathcal{X}_1|,|\mathcal{X}_2|)}$.
The quantum commuting family
$C_{qc}(|\mathcal{X}_1|,|\mathcal{X}_2|)$
is a closed set as well.
We note that if the systems were classical, then the three classes would have been identical based on the Fenchel-Eggleston-Carath\'eodory theorem \cite{Eggleston:66p} (see Section~\ref{subsection:Cardinality_Proof}). However, in the quantum case, the relation between those classes is not so obvious.

There are different conjectures related to the Tsirelson problem \cite{Tsirelson:93p};
the strong conjecture states that the commuting and spatial families are identical, i.e.,
$C_{qs}=C_{qc}$; the weak version that the commuting and approximate families are identical, i.e.,
$C_{qc}=C_{qa}$; and the middle states that the spatial and approximate families are identical, i.e., $C_{qs}=C_{qa}$. The weak Tsirelson conjecure is equivalent to Conne's embedding conjecture \cite{Ozawa:13p}.
The  Tsirelson conjecture is closely related to the study of nonlocal quantum   games:
$C_{qs}\neq C_{qa}$ is true if and only if there exists a nonlocal game that can be won with certainty using a limit of finite-dimensional entanglement resources, but cannot be won with certainty for a bounded dimension.

Slofstra \cite{Slofstra:19p} has shown that indeed there exists a game that can only be won in the limit of finite-dimensional entangelemnt, hence $C_{qs}$ is not closed, and both the strong and middle Tsirelson conjectures are false ($C_{qs}\neq C_{qc}$ and $C_{qs}\neq C_{qa}$).
More recently, Ji et al.  \cite{JNVWY:21p} constructed 
a game such that the entangled value differs from
its commuting-operator value, proving that the weak Tsirelson conjecture is also false.

As the dimension of our entangled systems $A_1$ and $A_2$ in the rate formulas (\ref{eq:inRetx_In}) and (\ref{eq:inRetx_Out_1}) is unbounded, the formulas for
$\mathcal{R}_{\text{ET}}(P_{Y|X_1,X_2})$ and $\mathcal{O}_{\text{ET}}(P_{Y|X_1,X_2})$ involve a union over  the approximate family
$C_{qa}$. 
Since $C_{qs}\neq C_{qa}$, it follows that the union cannot necessarily be exhausted by finite-dimensional bipartite states.

\begin{table*}\normalsize
\caption{MAC Settings with Entanglement and Conferencing Transmitters}
\label{Table:Conferencing}
\centering
\begin{tabular}{ |c|p{2cm}|p{2cm}|p{2cm}|p{2cm}||p{3cm}| }
 \cline{2-5}
 \multicolumn{1}{c}{}
&\multicolumn{3}{|c}{ {\textbf{Conferencing Stage}}}
&\multicolumn{1}{|c|}{{\textbf{Encoding Stage}}}
&\multicolumn{1}{c}{}\\
 \hline
 Case
& \multicolumn{1}{c|}{Classical Link} 
& \multicolumn{1}{c|}{Quantum Link}
& \multicolumn{1}{c|}{Entanglement?}
& \multicolumn{1}{c|}{Entanglement?}
&\multicolumn{1}{c|}{Capacity Formula}
\\
\hline
 \hypertarget{Case1}{\textcolor{blue!80!black}{1}}
&   %
\multicolumn{3}{c|}{\cellcolor{gray!5!}  No Conferencing}%
 &  \multicolumn{1}{c|}{\cellcolor{red!5!} No} & \multicolumn{1}{c|}{ \cite[Chap. 14]{CsiszarKorner:82b} }
\\
\hline
 \hypertarget{Case2}{\textcolor{blue!80!black}{2}}
&   \multicolumn{3}{c|}{\cellcolor{gray!5!} No Conferencing}
 &  \multicolumn{1}{c|}{\cellcolor{green!5!} Yes} & \multicolumn{1}{c|}{ $\mathcal{R}_{\text{ET}}$ (see (\ref{eq:inRetx_In})) }
\\
\hline
\hypertarget{Case3}{\textcolor{blue!80!black}{3}}
&   \multicolumn{1}{c|}{\cellcolor{red!5!}  \checkmark} &  
& \multicolumn{1}{c|}{\cellcolor{red!5!} No} &  \multicolumn{1}{c|}{\cellcolor{red!5!} No} & \multicolumn{1}{c|}{ \cite{Willems:83p} }
\\
 \hline
\hypertarget{Case4}{\textcolor{blue!80!black}{4}}
&   \multicolumn{1}{c|}{\cellcolor{red!5!}  \checkmark} &  
& \multicolumn{1}{c|}{\cellcolor{red!5!} No} &  \multicolumn{1}{c|}{\cellcolor{green!5!} Yes} & \multicolumn{1}{c|}{ $\mathcal{R}_{\text{ET-C}}$ (see (\ref{eq:inRetxD_Conferencing}))  }
\\
 \hline
 \hypertarget{Case5}{\textcolor{blue!80!black}{5}}
&   \multicolumn{1}{c|}{\cellcolor{red!5!}  \checkmark} &  
& \multicolumn{1}{c|}{\cellcolor{green!5!} Yes} &  \multicolumn{1}{c|}{\cellcolor{red!5!} No} & \multicolumn{1}{c|}{Open}
\\
 \hline
\hypertarget{Case6}{\textcolor{blue!80!black}{6}}
&   \multicolumn{1}{c|}{\cellcolor{red!5!}  \checkmark} &  
& \multicolumn{1}{c|}{\cellcolor{green!5!} Yes} &  \multicolumn{1}{c|}{\cellcolor{green!5!} Yes} & \multicolumn{1}{c|}{ $\mathcal{R}_{\text{ET-C}}$ (see (\ref{eq:inRetxD_Conferencing}))  }
\\
 \hline
\hypertarget{Case7}{\textcolor{blue!80!black}{7}}
&    &  \multicolumn{1}{c|}{\cellcolor{green!5!}  \checkmark}
& \multicolumn{1}{c|}{\cellcolor{red!5!} No} &  \multicolumn{1}{c|}{\cellcolor{red!5!} No} & \multicolumn{1}{c|}{ Open }
\\
 \hline
\hypertarget{Case8}{\textcolor{blue!80!black}{8}}
&    &  \multicolumn{1}{c|}{\cellcolor{green!5!} \checkmark}
& \multicolumn{1}{c|}{\cellcolor{red!5!} No} &  \multicolumn{1}{c|}{\cellcolor{green!5!} Yes} & \multicolumn{1}{c|}{ Open }
\\
 \hline
\hypertarget{Case9}{\textcolor{blue!80!black}{9}}
&    &  \multicolumn{1}{c|}{\cellcolor{green!5!}  \checkmark}
& \multicolumn{1}{c|}{\cellcolor{green!5!} Yes} &  \multicolumn{1}{c|}{\cellcolor{red!5!} No} & \multicolumn{1}{c|}{ Open }
\\
 \hline
\hypertarget{Case10}{\textcolor{blue!80!black}{10}}
&    &  \multicolumn{1}{c|}{\cellcolor{green!5!}  \checkmark}
& \multicolumn{1}{c|}{\cellcolor{green!5!} Yes} &  \multicolumn{1}{c|}{\cellcolor{green!5!} Yes} & \multicolumn{1}{c|}{ Open }
\\
 \hline
\end{tabular}

\end{table*}

\subsection{Open Problems}
\label{Subsection:Open}
There are many variations of the MAC setting with entangled and conferencing settings, as illustrated in Table~\ref{Table:Conferencing}. In particular, the conferencing transmitters may send either classical or quantum messages to one another, the conferencing protocol may consume pre-shared entanglement, and the encoding procedure may also use pre-shared entanglement, in addition to the conferencing outcome.  We refer to each case in Table~\ref{Table:Conferencing} below:
\begin{itemize}
\item
Case~\hyperlink{Case1}{\textcolor{blue!80!black}{1}} is the classical setting, without cooperation resources (see \cite[Chap. 14]{CsiszarKorner:82b}). The classical setting of Case~\hyperlink{Case3}{\textcolor{blue!80!black}{3}}, i.e., classical conferencing without entanglement, was originally studied by Willems \cite{Willems:83p}.

\item
Case~\hyperlink{Case1}{\textcolor{blue!80!black}{2}} corresponds to our first result in this work, on the MAC with entanglement resources between the transmitters, without conferencing. We have established in Theorem~\ref{theo:etMAC_In} that the capacity region is given by the regularization of the region $%
\mathcal{R}_{\text{ET}}(P_{Y|X_1,X_2})$.

\item
In the present work, we have introduced the setting in Case~\hyperlink{Case4}{\textcolor{blue!80!black}{4}}, whereby the transmitters perform a classical conferencing protocol and then encode using pre-shared entanglement. See Subsection~\ref{sec:Coding_Conferencing}. We have established in Theorem~\ref{theo:etMAC_Conferencing} that the capacity region is given by the regularization of the region $%
\mathcal{R}_{\text{ET-C}}(P_{Y|X_1,X_2},C_{12},C_{21})$.

\item
In Case~\hyperlink{Case1}{\textcolor{blue!80!black}{5}}, entanglement is only available at the conferencing stage.
In the encoding stage, the transmitters do not use entanglement resources.
Nevertheless, the transmitters can perform \emph{teleportation} in the conferencing stage in order to generate entanglement resources to be used in the encoding stage. Although, this could come at the expense of trading information between the transmitters.

\item
In Case~\hyperlink{Case6}{\textcolor{blue!80!black}{6}}, the transmitters may use pre-shared entanglement in the encoding stage as well.
As the conferencing link is classical and free of noise,  entanglement resources 
at the conferencing stage do not increase the achievable rates, 
and the capacity region in Case~\hyperlink{Case6}{\textcolor{blue!80!black}{6}} is the same as in Case~\hyperlink{Case4}{\textcolor{blue!80!black}{4}}.

\item 
The capacity region in the quantum conferencing settings remains open.
In Subsection~\ref{sec:Coding_Conferencing_Q},
we introduced the setting in Case~\hyperlink{Case10}{\textcolor{blue!80!black}{10}}, whereby the transmitters have quantum conferencing links, and they use pre-shared entanglement in both the conferencing and encoding stages.  We have shown that  the transmitters can then use the entanglement in order to double the conferencing rate using superdense coding  (see Theorem~\ref{theo:etMAC_Conferencing_Q}).

\item
We conjecture that the capacity region in Case~\hyperlink{Case8}{\textcolor{blue!80!black}{8}} is the same as in Case~\hyperlink{Case4}{\textcolor{blue!80!black}{4}}, i.e., if the transmitters have unlimited entanglement resources at the encoding stage, but none at the conferencing stage, then it makes no difference whether the conferecing link is classical or quantum. 

\item
Notice that in Case~\hyperlink{Case7}{\textcolor{blue!80!black}{7}}, the encoders do not have access to entanglement resources.
Nevertheless, quantum conferencing can be used in order to \emph{generate} entanglement resources. Yet, this could come at the expense of trading information between the transmitters. The same principle holds in Case~\hyperlink{Case9}{\textcolor{blue!80!black}{9}}. The optimum in those cases remains open as well.

\end{itemize}

\section*{Acknowledgment}
The authors wish to thank Andreas Winter (Universitat Aut\`onoma de Barcelona) for useful discussions and important observations on the dimension problem.

U. Pereg was supported by the
 Israel Science Foundation (ISF), Grants 939/23 and 2691/23, German-Israeli Project Cooperation (DIP) within the  Deutsche Forschungsgemeinschaft (DFG), Grant
2032991, 
the Junior Faculty Program for Quantum Science and Technology of the Israel Planning and Budgeting Committee of the Council for Higher Education   (VATAT)  through Grant 86636903, Chaya Career Advancement Chair, Grant 8776026, and the Nevet Program of the Helen Diller Quantum Center at the Technion, Grant 2033613.
C. Deppe and H. Boche acknowledge the financial support by the Federal
Ministry of Education and Research (BMBF) of Germany in the programme of ``Souver\"an. Digital. Vernetzt." Joint project 6G-life, project identification number: 16KISK002.
H. Boche was supported in part by the BMBF through the grants 16KISQ093 (QUIET), 16KIS1598K
(QuaPhySI), 16KISQ077(QDCamNetz) and 16KISR027K (QTREX).

\begin{appendix}[Maximal Error Criterion]
We show here that if a rate pair $(R_1,R_2)$ is achievable with  a 
message-average error criterion, then it is also achievable with a maximal error criterion (see (\ref{Equation:Message_Max_Error}), \cf  (\ref{Equation:Message_Avg_Error})). 
As pointed out in Remark~\ref{Remark:Classical_Maximal_Error}, this property 
does \emph{not} hold for a classical MAC without entanglement resources
\cite{Dueck:78p} \cite[Sec. 2.2]{Cai:14p}.
Nonetheless, Cai \cite{Cai:14p} considered the classical MAC without entanglement resources, and showed that
 the capacity region of the classical MAC  with a message-average error criterion is also achievable with a maximal error criterion, if the encoders are provided with a random key. Since the entanglement resources in our model can also be used in order to generate a random key, the arguments by 
 Cai \cite{Cai:14p} extend to our model as well. For completeness, we give the details below.
 It can easily be seen from the calculations below that the new codes only require rate-constrained entanglement resources.

We note that if $(R_1,R_2)=(0,0)$ is the best  achievable pair with a message-average error criterion, then the capacity region with a maximal error criterion is  $\mathcal{C}_{\text{ET}}(P_{Y|X_1,X_2})=\{(0,0)\}$, and there is nothing to show. Hence, we may assume without loss of generality that $R_1>0$ is achievable with entangled transmitters subject to a message-average error criterion.
 Let $\mathscr{C}%
 $ be a code with entangled transmitters such that the message-average error probability is bounded by
 \begin{align}
&\overline{P}_{e}^{(n)}(\mathscr{C})\equiv %
\frac{1}{2^{n(R_1+R_2)}}\sum_{m_1=1}^{2^{nR_1}}\sum_{m_2=1}^{2^{nR_2}}    P_{e}^{(n)}(\mathscr{C}|m_1,m_2)< \gamma %
\label{Equation:Message_Avg_Error_App}
\end{align}
for large $n$,
where $\gamma$ %
is arbitrarily small.
For every $m_2\in [1:2^{nR_2}]$, define the semi-average probability of error by
 \begin{align}
&e(m_2)\equiv %
\frac{1}{2^{nR_1}}\sum_{m_1=1}^{2^{nR_1}}    P_{e}^{(n)}(\mathscr{C}|m_1,m_2) 
\,.
\label{Equation:Message_1_Avg_Error}
\end{align}

Consider the following subset of ``good" messages for User 2,
\begin{align}
\Mset_{2}'&=\left\{ m_2\in [1:2^{nR_2}] \,:\; e(m_2)< 2\cdot\gamma
\right\} \,.
\label{Equation:Mset_2_p}
\end{align}
Observe that by (\ref{Equation:Message_Avg_Error_App}), the average value of 
$e(m_2)$ is bounded by
\begin{align}
\frac{1}{2^{nR_2}}\sum_{m_2=1}^{2^{nR_2}}    e(m_2)< 
\gamma
\,.
\label{Equation:Average_e_m2}
\end{align}
This, in turn, implies that the subset
$\Mset_{2}'$ is at least as large as half the original set, i.e.,
\begin{align}
|\Mset_{2}'|&\geq \frac{1}{2}\cdot 2^{nR_2}
\\
&= 2^{n\left( R_2-\frac{1}{n} \right)}
\,.
\end{align}
Otherwise, the average value of $e(m_2)$ would have been greater than or equal to 
$\gamma$, %
in contradiction to (\ref{Equation:Average_e_m2}).
Therefore, \mbox{User 2} can throw away the messages outside $\Mset_{2}'$, and transmit at a rate of $R_2'=R_2-\frac{1}{n}$, arbitrarily close to $R_2$.

We now construct a new code that is reliable under the maximal error criterion.
In this construction, Encoder 1 first performs a measurement on an ancilla of dimension $n^2$ to obtain a uniformly distributed key $L_1\in [1:n^2]$. 
Then, the transmission consists of two consecutive blocks.
In the first block, Encoder 1 sends the key $L_1$ to the receiver using a code based on a \emph{message-average} error criterion. At the same time, Encoder 2 sends $L_2=1$.
Since the key set size is sub-exponential and we have assumed that User 1 can communicate at a positive rate, the key can be sent using a code of length $\nu=o(n)$.
As for the second block,
given a key outcome $L_1$, Encoder 1 chooses a permutation $\pi_{L_1}$
on the message set $[1:2^{nR_1}]$, and encodes by applying the encoding map
$\, \Fset_1^{(\tilde{m}_1)}$ with
$\tilde{m}_1\equiv \pi_{L_1}(m_1)$. At the same time, Encoder 2  encodes using 
$\Fset_2^{(m_2)}$, as in the original code.
Bob receives the output sequence $(\bar{Y}^{\nu},Y^n)$ of length $\nu+n$, and decodes as follows. He uses the first part, $\bar{Y}^{\nu}$, in order to find an estimate  $\hat{L}_1$ and $\hat{L}_2=1$ for the keys.
Then, Bob declares his estimation for the messages as 
$(\hat{m}_1,\hat{m_2})=(\pi_{\hat{L}_1}^{-1}\times \identity_2)g(Y^n)$,
using the output $Y^n$ of the second block,
 where $\identity_2$ is the identity permutation on $\Mset_2'$.
 We denote the code that is used in the second block by $\pi_L(\mathscr{C})=(\Psi,\pi\Fset_1,\Fset_2,\pi^{-1}g)$.

Having assumed that the key $L_1$ is uniformly distributed, the probability of decoding the wrong key is the semi-average probability of error for the first block. That is, 
\begin{align}
\Pr(\hat{L}_1\neq L_1)=
\frac{1}{n^2} \sum_{\ell=1}^{n^2}
P_{e}^{(\nu)}(\mathscr{C}_1|\ell,1)
< 2\gamma %
\,.
\end{align}

Now, consider the second block.
Let $\mathfrak{P}$ denote the permutation group  on the message set $[1:2^{nR_1}]$ of User 1.
Furthermore, let $\Pi_1,\ldots,\Pi_{n^2}$ be an i.i.d. sequence of random permutations, each uniformly distributed over $\mathfrak{P}$. 
Then, for a given $m_1$,
\begin{align}
\Pr(\Pi_\ell(m_1)=\tilde{m}_1)=\frac{ (2^{nR_1}-1)! }{(2^{nR_1})!}=\frac{1}{2^{nR_1}}
\label{Equation:Permutation_Distribution}
\end{align}
for all
$\tilde{m}_1\in [1:2^{nR_1}]$ and $\ell\in [1:n^2]$.
Thus, for every message pair $(m_1,m_2)\in [1:2^{nR_1}]\times \Mset_2'$,
\begin{align}
\mathbb{E}\left[ P_e^{(n)}(\Pi_\ell(\mathscr{C})|m_1,m_2) \right]
&=\sum_{\tilde{m}_1} \Pr(\Pi_\ell(m_1)=\tilde{m}_1) \cdot
P_e^{(n)}(\mathscr{C}| \tilde{m}_1,m_2 )
\\
&=\frac{1}{2^{nR_1}} \sum_{\tilde{m}_1} 
P_e^{(n)}(\mathscr{C}| \tilde{m}_1,m_2 )
\\
&= e(m_2)
\\
&<2\gamma %
\end{align}
where the first equality follows from the construction of the code
$\pi_\ell(\mathscr{C})$, the second equality is due to (\ref{Equation:Permutation_Distribution}),
the third is due to (\ref{Equation:Message_1_Avg_Error}), and the last inequality 
follows from the definition of $\Mset_2'$ in (\ref{Equation:Mset_2_p}).
Hence, based on the Chernoff bound, we have
\begin{align}
\Pr\left( \frac{1}{n^2}\sum_{\ell=1}^{n^2}  P_e^{(n)}(\Pi_\ell(\mathscr{C})|m_1,m_2)
> 7\gamma
\right)\leq e^{-\gamma\cdot n^2}
\end{align}
(see \cite[Lemma 3.1]{Cai:14p}, taking $L\leftarrow n^2$, $\alpha\leftarrow 2\gamma$,
$\beta\leftarrow 7\gamma$).
This means that the probability that the random code $\Pi_L(\mathscr{C})$ has an error higher than 
$7\gamma$
tends to zero in a super-exponential rate.
Thus, by the union bound,
\begin{align}
\Pr\left(\exists (m_1,m_2) \,:\; \frac{1}{n^2}\sum_{\ell=1}^{n^2}  P_e^{(n)}(\Pi(\mathscr{C})|m_1,m_2)
> 7\gamma
\right)&\leq 2^{n(R_1+R_2')}\cdot e^{-\gamma\cdot n^2} 
\\
&\leq  e^{-\frac{1}{2}\gamma\cdot n^2} 
\end{align}
for sufficiently large $n$.
Therefore, there exists a realization $(\pi_1,\ldots,\pi_{n^2})$, such that
\begin{align}
P_e^{(n)}(\pi_L(\mathscr{C})|m_1,m_2)=
\frac{1}{n^2}\sum_{\ell=1}^{n^2}  P_e^{(n)}(\pi_\ell(\mathscr{C})|m_1,m_2)
\leq 7\gamma
\end{align}
for all $(m_1,m_2)\in [1:2^{nR_1}]\times\Mset_2'$.
In other words, the maximal error probability for the second block is bounded by $7\gamma$.
By choosing $\gamma=\frac{1}{9}\varepsilon_0$, we have that the maximal error probability of the overall code of length $n+o(n)$ %
is bounded by $\varepsilon_0$, as we wanted to show.
By increasing the code length $n$, 
 the overall transmission rates
$\frac{n}{n+o(n)}R_1$ and $\frac{n}{n+o(n)}\left(R_2-\frac{1}{n}\right)$ can be made arbitrarily close to
 $R_1$ and $R_2$, respectively. \qed

\end{appendix}

\bibliography{References}{}

\end{document}